# Emergent photons and fractionalized excitations in a quantum spin liquid

Bin Gao[1,*], Félix Desrochers[2,*], David W. Tam[3,*], Paul Steffens[4], Arno Hiess[4,5], Yixi Su[6], Sang-Wook Cheong[7], Yong Baek Kim[2], and Pengcheng Dai[1]

[1] Department of Physics and Astronomy, Rice University, Houston, Texas 77005, USA

[2] Department of Physics, University of Toronto, Toronto, Ontario M5S 1A7, Canada

[3] Laboratory for Neutron Scattering and Imaging, Paul Scherrer Institut, CH-5232 Villigen PSI, Switzerland

[4] Institut Laue-Langevin, 71 Avenue des Martyrs, 38042 Grenoble Cedex 9, France

[5] European Spallation Source ERIC, 22100 Lund, Sweden

[6] Jülich Centre for Neutron Science (JCNS) at Heinz Maier-Leibnitz Zentrum (MLZ), Forschungszentrum Jülich, Lichtenbergstrasse 1, D-85747 Garching, Germany

[7] Rutgers Center for Emergent Materials and Department of Physics and Astronomy, Rutgers University, Piscataway, New Jersey 08854, USA

[*]These authors made equal contributions to this work

**A quantum spin liquid (QSL) arises from a highly entangled superposition of many degenerate classical ground states in a frustrated magnet, and is characterized by emergent gauge fields and deconfined fractionalized excitations (spinons)[1-5]. Because such a novel phase of matter is relevant to high-transition-temperature superconductivity[6,7] and quantum computation[8,9], the microscopic understanding of QSL states is a long-sought goal in condensed matter physics. The 3D pyrochlore**

**lattice of corner-sharing tetrahedra (Fig. 1a) can host a QSL with $U(1)$ gauge fields called quantum spin ice (QSI)[10], which is a quantum (with effective $S = 1/2$) analog of the classical (with large effective moment) spin ice[11-13]. A key difference between QSI and classical spin ice is the predicted presence of the linearly dispersing collective excitations near zero energy, dubbed the "photons", arising from emergent quantum electrodynamics, in addition to the spinons at higher energies[10,14,15]. Recently, 3D pyrochlore systems $Ce_2M_2O_7$ ($M$ = Sn, Zr, Hf) have been suggested as effective $S = 1/2$ QSI candidates[16-23], but there has been no evidence of quasielastic magnetic scattering signals from photons, a key signature for a QSI. Here, we use polarized neutron scattering experiments on single crystals of $Ce_2Zr_2O_7$ to conclusively demonstrate the presence of magnetic excitations near zero energy at 50 mK in addition to signatures of spinons at higher energies. By comparing the energy ($E$), wave vector ($Q$), and polarization dependence of the magnetic excitations with theoretical calculations[24-30], we conclude that $Ce_2Zr_2O_7$ is the first example of a dipolar-octupolar $\pi$-flux QSI[28] with dominant dipolar Ising interactions, therefore identifying a microscopic Hamiltonian responsible for a QSL.**

QSLs are novel phases of interacting quantum spins ($S = 1/2$) in a crystalline solid that have long-range entangled ground states and no magnetic order down to zero temperature. Although Anderson first proposed their existence for 2D triangular lattice some 50 years ago[31], conclusive identification of a QSL material and its associated microscopic Hamiltonian is still lacking[1-5]. For the $S = 1/2$ 2D honeycomb lattice, Kitaev's exactly solvable model with bond-dependent nearest-neighbor interactions has a QSL ground state, where the excitations are itinerant Majorana fermions and static $Z_2$ fluxes relevant for fault-tolerant quantum computation[8,9]. Despite intensive efforts, there is currently no conclusive identification of a Kitaev QSL material[32]. For $S = 1/2$ 2D kagome and triangular lattice magnets, although there are many reports of fractionalized excitations consistent with spinon Fermi surface QSL[33-40], the microscopic Hamiltonian are difficult to simulate which makes it challenging to formulate any concrete prediction[34,41,42].

In 3D rare-earth pyrochlore magnets with large effective moment, Ising-like spins decorating the corner-sharing tetrahedra (Fig. 1a) form a constrained paramagnet where the system is energetically restricted to the degenerate "2-in-2-out" classical spin ice (CSI) states, analogous to the "2-near-2-far" rules of the covalent $2H^+$-$O^{2-}$ bonding distances in water ice[43,44]. This set of local constraints, known as the ice rules, is imposed on every tetrahedron and can be mapped to a divergence-less coarse-grained spin field $\mathcal{E}$. Defect tetrahedra where the local ice rules are violated then behave as mobile charged excitations interacting electrostatically through this emergent field. This structure leads to highly distinctive signatures in neutron scattering. For single crystals such as $Ho_2Ti_2O_7$ aligned in the $(h, h, 0) \times (0,0, l)$ scattering plane (Fig. 1b), CSI has "pinch points" in the $\boldsymbol{Q}$ dependence of the spin-spin correlation function $S(\boldsymbol{Q})$ for the magnetic fluctuations $M_y$ in the scattering plane perpendicular to the momentum transfer (Fig. 1c). The pinch points can be revealed in polarized neutron scattering experiments with neutrons polarized along the $z$-axis perpendicular to the scattering plane (Fig. 1c)[12,43,44]. These seminal results showcase the usefulness of polarized neutron scattering in unveiling exotic emergent physics and their connection to microscopic descriptions.

For $S = 1/2$ pyrochlore magnets where quantum effects are important, CSI is theoretically predicted to be promoted to QSI, a QSL where the low-energy dynamics of the 2-in-2-out manifold is described by compact quantum electrodynamics in analogy to the $U(1)$ gauge structure of Maxwell's equations[14,15,45-49]. This implies that QSI hosts gapless excitations, dubbed photons, with a linear dispersion that describes coherent fluctuations within the spin ice manifold (Fig. 1e). QSI also supports massive $S = 1/2$ spinon excitations that are the quantum analog of the defect tetrahedra in CSI (Fig. 1f) and gapped topological defects known as visons[10]. The experimental discovery of a material realization of QSI is an outstanding problem that, despite encouraging results[50-53], has yet to receive any definitive evidence.

**Experimental and Theoretical Results**

Recently, Ce-based dipolar-octupolar pyrochlores $Ce_2M_2O_7$ (*M*=Sn, Zr, Hf) have been proposed as QSI candidates[16-23]. The $Ce^{3+}$ ($4f^1$, $^2F_{5/2}$) magnetically active ions in the crystal field of eight oxygen anions form an effective $S = 1/2$ Kramers doublet. The doublets can be modeled as pseudospins-½ where the x- and z-components of the pseudospin ($\tau^x, \tau^z$) transform as magnetic dipoles while the y-component ($\tau^y$) transforms as a magnetic octupole[24,25]. The x-component is referred to as a dipole despite having an octupolar magnetic charge density (Fig. 1a) because of its transformation property. The most general nearest-neighbor Hamiltonian for these dipolar-octupolar pseudospin systems is

$$\mathcal{H} = \sum_{a \in \{x,y,z\}} \sum_{\langle i,j \rangle} J_{aa} \tau_i^a \tau_j^a + \sum_{\langle i,j \rangle} J_{xz}(\tau_i^x \tau_j^z + \tau_i^z \tau_j^x), \qquad (1)$$

where the pseudospin components are defined in sublattice-dependent local coordinates. By performing a local rotation about the y-axis, the system can be brought to the simple XYZ form

$$\mathcal{H} = \sum_{a \in \{x,y,z\}} \sum_{\langle i,j \rangle} \mathcal{J}_a S_i^a S_j^a, \qquad (2)$$

with the pseudospins $S^x = \cos\theta\, \tau^x - \sin\theta\, \tau^z$, $S^y = \tau^y$, and $S^z = \sin\theta\, \tau^x + \cos\theta\, \tau^z$. When the dipolar-octupolar system stabilizes QSI, the XYZ model can be mapped to lattice quantum electrodynamics by associating the pseudospin component with the dominant coupling $\mathcal{J}_\| = max(\mathcal{J}_x, \mathcal{J}_y, \mathcal{J}_z)$ to the emergent electric field $\mathcal{E}$, $S_i^\| = \mathcal{E}_{r,r'}$, and the transverse parts to spinon bilinears dressed with the emergent photon $S_i^+ = \frac{1}{2}\Phi_r^\dagger e^{i\mathcal{A}_{r,r'}}\Phi_{r'}$. Here $S^+$ is the raising operator in the $S^\|$ basis, $r$ and $r'$ label the centers of the tetrahedron such that the site $i$ sits at the center of the link $r \to r'$ joining them, $\Phi_r^\dagger$ is a spinon creation operator, and $\mathcal{A}$ is the vector potential canonically conjugate to the $\mathcal{E}$ field. The interpretation of the above construction is that $S^\|$ corresponds to the emergent electric field, which is divergence-less (i.e., $\nabla \cdot \mathcal{E} = 0$) in the 2-in-2-out manifold (i.e. when $\mathcal{J}_\|$ is much larger than the other couplings). Flipping this lattice field with $e^{i\mathcal{A}_{r,r'}}$ breaks the ice rules (i.e., $\nabla \cdot \mathcal{E} \neq 0$) and must then be accompanied by the creation of a spinon-antispinon pair (Fig. 1e) that act as sources of the emergent field.

Depending on the leading coupling $\mathcal{J}_\|$, drastically different responses are expected in inelastic neutron scattering. Indeed, Ce$_2$Zr$_2$O$_7$ has an anisotropic $g$-tensor with $g_{zz} = 2.57$ and $g_{xx} = g_{yy} = 0$ such that the magnetic field linearly couples only to $\tau^z$ [16,17]. As a result, neutron scattering cross sections σ are only sensitive to correlations between $\tau^z$ for small momentum transfer $\sigma \sim \langle \tau^z \tau^z \rangle \sim \cos^2\theta \langle S^z S^z \rangle + \sin^2\theta \langle S^x S^x \rangle$ [19]. If $\mathcal{J}_\parallel = \mathcal{J}_y > \mathcal{J}_x, \mathcal{J}_z$, inelastic neutron scattering probes the two-spinon continuum $\sigma \sim \cos^2(\theta)\langle \Phi^\dagger \Phi \Phi^\dagger \Phi \rangle + \sin^2(\theta)\langle \Phi^\dagger \Phi \Phi^\dagger \Phi \rangle$ whereas the photon remains invisible. In contrast, for $\mathcal{J}_\parallel = \mathcal{J}_x > \mathcal{J}_y, \mathcal{J}_z$, both the photons and spinons should be observable provided there is non-zero off-diagonal term $\theta \neq 0$ (i.e., $J_{xz} \neq 0$) since $\sigma \sim \cos^2(\theta)\langle \Phi^\dagger \Phi \Phi^\dagger \Phi \rangle + \sin^2(\theta)\langle \mathcal{E}\mathcal{E} \rangle$. Identifying a quasielastic signal from the photons thus provides a method to obtain crucial information about the underlying microscopic couplings of a dipolar-octupolar QSI (Fig. 1d).

Previous investigations of Ce$_2$Zr$_2$O$_7$ have strongly hinted that it stabilizes a QSI as its ground state. Heat capacity, magnetic susceptibility, and muon spin relaxation find no sign of magnetic order or spin freezing[16,17,20,54-56]. Further detailed theoretical fits to thermodynamic measurements have determined the microscopic couplings. The current consensus is that the system is in a region of parameter space that is theoretically suggested to host a specific flavor of QSI known as the π-flux quantum spin ice (π-QSI), where a static π-flux of the emergent gauge field threads the hexagonal plaquette of the pyrochlore lattice ($\nabla \times \mathcal{A} = \pi$). The other stable QSL in the phase diagram, 0-flux QSI (0-QSI) (see Fig. 1d), only has vanishing fluxes ($\nabla \times \mathcal{A} = 0$). In particular, one investigation found that the leading coupling is between the octupolar components $S^y$ [27], whereas another puts Ce$_2$Zr$_2$O$_7$ at the boundary (i.e., $\mathcal{J}_y \approx \mathcal{J}_x > \mathcal{J}_z$) between π-flux QSI with dominant dipolar (π-D-QSI) and dominant octupolar (π-O-QSI) coupling (Fig. 1d)[55]. As explained above, inelastic neutron scattering should be able to differentiate between these two cases since the photon will not be visible in π-O-QSI but should produce a visible quasielastic signal for π-D-QSI if θ ≠ 0.

In previous experiments on $Ce_2Zr_2O_7$ and $Ce_2Sn_2O_7$, unpolarized neutron scattering or polarized neutron scattering with only one neutron polarization direction were carried out at base ($T = 35 - 100$ mK) and high ($T \approx 10$ K) temperatures. Then the magnetic signal was extracted by taking the temperature difference in scattering signals between the base and high temperature[16-20,54,55,57]. This method reveals a broad continuum consistent with a two-spinon continuum and no magnetic signal at the elastic position $E \approx 0$ meV. In particular, recent measurements on $Ce_2Sn_2O_7$ have revealed the presence of three inelastic peaks of decreasing intensity in the continuum[57], a signature of π-flux QSI (Fig. 1d)[30]. Although such measurements are consistent with π-O-QSI, determining the existence/absence of a quasielastic magnetic signal is an involved task that requires careful consideration. The fundamental assumption in isolating the magnetic signal by temperature subtraction is that the nonmagnetic scattering is temperature-independent between the base and high temperature[16-20,54,55,57]. Although such an assumption might be reasonable for CSI where magnetic scattering is much larger than the nonmagnetic contributions due to the large effective moment[12,43,44], it is unclear that the temperature difference method[16-20,54,55,57] can effectively extract the magnetic signal in the low-energy region, near $E \approx 0$, for a $S = 1/2$ system with a potentially large nonmagnetic scattering background.

A conclusive way to resolve this problem and unambiguously isolate the energy ($E$) and momentum ($\boldsymbol{Q}$) dependence of magnetic scattering in a material is the full neutron polarization analysis[58-60]. By polarizing incident beam neutrons along the $x$-axis (parallel to $\boldsymbol{Q}$), $y$-axis (perpendicular to $\boldsymbol{Q}$ in the scattering plane), and $z$-axis direction (perpendicular to the scattering plane), the neutron spin-flip (SF) and non-spin-flip (NSF) scattering cross sections at $\boldsymbol{Q}$ and $E$ for $x, y,$ or $z$ polarization direction are $\sigma_{x,y,z}^{SF}(\boldsymbol{Q}, E)$ and $\sigma_{x,y,z}^{NSF}(\boldsymbol{Q}, E)$, respectively (Fig. 1c). They are related to the magnetic scattering along the $y$ and $z$ directions $M_y$ and $M_z$ via[60]:

$$\begin{pmatrix} \sigma_x^{SF} \\ \sigma_y^{SF} \\ \sigma_z^{SF} \\ \sigma_x^{NSF} \\ \sigma_y^{NSF} \\ \sigma_z^{NSF} \end{pmatrix} = \frac{1}{R+1} \begin{pmatrix} R & R & 1 & (2R+1)/3 & (R+1) \\ 1 & R & 1 & (2R+1)/3 & (R+1) \\ R & 1 & 1 & (2R+1)/3 & (R+1) \\ 1 & 1 & R & (R+2)/3 & (R+1) \\ R & 1 & R & (R+2)/3 & (R+1) \\ 1 & R & R & (R+2)/3 & (R+1) \end{pmatrix} \begin{pmatrix} M_y \\ M_z \\ N \\ NSI \\ B \end{pmatrix}, \qquad (3)$$

where $R$ is the imperfect neutron polarization quantified by measuring the nuclear Bragg peaks contamination into the SF channel ($R = \frac{NSF_N}{SF_N} \approx 30$ in our experimental set up), $N$ is nuclear coherent scattering (including temperature-dependent phonon scattering), $NSI$ is the $\boldsymbol{Q}$-independent nuclear spin incoherent scattering ($NSI \ll N$), and $B$ is the background scattering. For polarized neutron scattering experiments measuring only $\sigma_z^{SF}(\boldsymbol{Q})$ and $\sigma_z^{NSF}(\boldsymbol{Q})$[12,20], both channels would have significant $NSI + B$ scattering, and the $\sigma_z^{NSF}(\boldsymbol{Q})$ would also have nearly full $N$ contribution. For CSI systems such as Ho$_2$Ti$_2$O$_7$ with large local moments ($M_y, M_z \gg N, NSI, B$), one can approximate $\sigma_z^{SF}(\boldsymbol{Q}) \approx M_y$ and $\sigma_z^{NSF}(\boldsymbol{Q}) \approx M_z$[12]. However, for the effective $S = 1/2$ system such as Ce$_2$Zr$_2$O$_7$, there is no determination of the intensity ratio between $M_y, M_z$ and $N, NSI, B$ at different temperatures[20]. Therefore, full neutron polarization analysis by measuring $\sigma_{x,y,z}^{SF}(\boldsymbol{Q},E)$ and $\sigma_x^{NSF}(\boldsymbol{Q},E)$ is necessary to conclusively separate the magnetic scattering from the $N, NSI, B$ without the need to use high-temperature measurement to estimate the background scattering[20].

Figure 2a shows energy scans of $\sigma_{x,y,z}^{SF}(\boldsymbol{Q},E)$ and $\sigma_x^{NSF}(\boldsymbol{Q},E)$ at $\boldsymbol{Q} = (0,0,1)$ and $T = 50$ mK. Near the elastic position ($E = 0 \pm 0.03$ meV), we find $\sigma_x^{NSF} > \sigma_x^{SF} > \sigma_y^{SF} \approx \sigma_z^{SF}$. At the inelastic position where previous unpolarized neutron scattering work found spin excitation continuum around $E = 0.12$ meV[16,17], we have $\sigma_x^{SF} > \sigma_y^{SF} \approx \sigma_z^{SF} > \sigma_x^{NSF}$. The energy dependence of $M_z + M_y$ determined using $\sigma_{x,y,z}^{SF}$ in Fig. 2a and Eq. (3) reveals three distinct peaks at $E \approx 0, 0.05, 0.12$ meV with dominating magnetic scattering at $E = 0 \pm 0.03$ meV (Figure 2d). Furthermore, $M_z - M_y \approx 0$ revealed in Fig. 2j indicates that the magnetic scattering at probed energies is isotropic in spin space. While the magnetic intensity around $E \approx 0.12$ meV is consistent with unpolarized neutron scattering measurements[16,17], the discovery of dominating magnetic scattering near zero energy is surprising and cannot be obtained from previous work[16-20,54,55,57]. Since unpolarized neutron scattering measures $\sigma_x^{NSF} + \sigma_x^{SF}$, we estimate that $M_z + M_y$ is about 10% and 75% of the total scattering at the elastic position and $E = 0.12$ meV, respectively (Figs. 2a and 2d). Figures 2b and 2c summarize energy scans of $\sigma_{x,y,z}^{SF}(\boldsymbol{Q},E)$ and $\sigma_x^{NSF}(\boldsymbol{Q},E)$ at $\boldsymbol{Q} = (3/4, 3/4, 0)$ and $(1,1,0)$, respectively. The estimated $M_z + M_y$ and $M_z - M_y$ are shown in Fig. 2e and 2k, respectively, for $\boldsymbol{Q} = (3/4, 3/4, 0)$. Similar results, obtained with better instrumental energy resolution, are shown in Figs. 2f and 2l. While magnetic scattering at

X and K points are isotropic in spin space at probed energies (Figs. 2j and 2k), the elastic magnetic scattering at $\boldsymbol{Q} = (1,1,0)$ is clearly anisotropic with $M_z - M_y > 0$ (Fig. 2l).

To understand these results, we model the spinon dynamics using the framework of gauge mean-field theory (GMFT)[29,30,61-64] and describe the photons using Gaussian quantum electrodynamics (see methods)[14,15]. In our modeling, the spinons and photons are effectively decoupled such that the spinons are only sensitive to the average background π-flux threading the hexagonal plaquettes. Assuming such an emergent quantum electrodynamics description, the dominant quasielastic signal is then explained by the emergent photons. We reach this conclusion since the dominant quasielastic signal cannot be accounted for by only invoking the spinon contribution (see methods). As already emphasized, observation of a signal coming from the emergent photons implies that $\mathcal{J}_x > \mathcal{J}_y, \mathcal{J}_z$ and $\theta \neq 0$ (π-D-QSI regime).

We fit the experimental results by tuning the microscopic couplings and the speed of the emergent photon to obtain the theoretical predictions reported in Fig. 2g-2l. The theoretical modeling predicts a series of peaks in the total magnetic scattering $M_z + M_y$. The first quasielastic one comes from photons, and then the spinons produce three others of decreasing intensity (the third is extremely faint). These three inelastic spinon peaks are a unique and distinctive signature of π-flux QSI[30]. In contrast, in 0-flux QSI, spinons produce a broad inelastic continuum with a single local maximum. The predicted transitions are seen in the experimental data as is especially clear at $\boldsymbol{Q} = (0,0,1)$ in Fig. 2d. At this point, there is a transition from the photon contributions to the first spinon peak at around $E \approx$

0.025 meV and then a second one from the first to the second spinon peak at $E \approx 0.075$ meV. Our model is consistent with the position of these transitions and the decreasing intensity of the corresponding peaks. For the polarization difference $M_z - M_y$, theory predicts that there should not be any anisotropy at $Q = (0,0,1)$, a small response at $Q = (3/4,3/4,0)$ and the most intense signal at $Q = (1,1,0)$ which is once again consistent with measurements (Fig. 2j-2l). In particular, at $Q = (1,1,0)$ in Fig. 2l, the anisotropic quasielastic response is well accounted for by the photon. The main discrepancy is that the model predicts a finite $M_z - M_y$ at $Q = (1,1,0)$ in the inelastic spinon contribution, which is not seen in the experiment. This could potentially be attributed to effects beyond GMFT, such as thermally excited fluxes or spinons-photon interactions. In short, the energy scans provide evidence for a substantial quasielastic signal with the same momentum-dependent polarization anisotropy as predicted for the emergent photon, as well as multiple inelastic peaks consistent with the spinon continuum of π-flux QSI.

To investigate further if the large quasielastic signal is compatible with the emergent photon, we examine its momentum dependence (Fig. 3). Fig. 3a and 3c present raw $\sigma_{x,y,z}^{SF}(Q)$ and the resulting polarization anisotropy $M_z - M_y$ at specified $Q$ with $E = 0 \pm 0.03$ meV. The quasielastic signal displays clear momentum and polarization dependence with $M_z > M_y$ at $Q = (1,1,0)$, inconsistent with conventional magnetic disorder where one would expect a polarization- and momentum-independent signal (i.e., $M_z \approx M_y$ at all $Q$) (see methods). In contrast, our modelling using emergent spinons and photons properly captures the observed spin space anisotropy (Fig. 3b-3c). Fig. 3d and 3h show the calculated $Q$-dependence of $M_y$ and $M_z$ at $E = 0 \pm 0.03$ meV, respectively. These are compared with measurements of $M_y$ in Fig. 3e-3g and of $M_z$ presented in Fig. 3i-3k. Both comparisons show that the GMFT and Gaussian quantum electrodynamics nicely reproduce the momentum and polarization dependence of the magnetic scattering. In particular, Fig. 3e and 3g show a depletion $M_y$ at the zone center and oscillations along the $[0,0,l]$ direction. These observations are often taken as signatures of QSI given that one expects a flat $M_y$ signal at the zone center and along $[0,0,l]$ for CSI (Fig. 3g)[12,15,16,43,44,52]. Since the magnetic scattering at the elastic line is dominated by the photon, as illustrated in Fig. 3i-3k, we fit these measurements to determine an optimal speed of light of $\hbar c_{QSI}/a_0 = 0.65\ k_B T_{exp.} = 0.0028$ meV, where $a_0$ is the lattice constant.

Fig. 4a and 4d show the calculated wave vector dependence of $M_y$ and $M_z$ at $E = 0.1 \pm 0.03$ meV in the $(h, h, l)$ zone, respectively. By comparison with Fig. 2, this momentum scan is centered around the second spinon peak and is thus dominated by spinons with only negligible photon scattering. The comparison of experimentally estimated $M_y$ and $M_z$ with theoretical calculations along the $[0,0,l]$ and $[h,h,0]$ directions are summarized in Figs. 4b-4c and Figs. 4e-4f. It should be emphasized that theoretical calculations are scaled consistently between Figs. 2, 3 and 4. These results thus show that, on top of giving a reasonable qualitative agreement for the polarization and momentum dependence of the inelastic magnetic scattering, our description in terms of emergent quantum electrodynamics quantitatively reproduces the intensity ratio between the elastic signal and the inelastic one at $E \approx 0.1$ meV.

**Summary**

We performed full polarization analysis of neutron scattering experiments on a 3D pyrochlore lattice QSI candidate $Ce_2Zr_2O_7$ and have discovered a continuum of quasielastic magnetic scattering around zero energy. This signal is incompatible with a spinon continuum at higher energy or magnetic disorder but consistent with the emergent photon predicted to exist in QSI[10,14,15]. Such an observation also sheds light on the microscopic couplings since the photon detection is only possible if the ground state is π-D-QSI, not π-O-QSI. We have further highlighted that the multiple peaks structure of the inelastic signal and the momentum dependence of the intensity profile all display key features of π-QSI. The above interpretation will certainly require future confirmation through higher-resolution measurements of the quasielastic signal and identification of its temperature dependence. However, this work combined with previous experimental investigations[16,17,20,54-56] lends compelling support to the identification of $Ce_2Zr_2O_7$ as an experimental realization of QSI — one of the most paradigmatic QSL in condensed matter physics.

**Methods:**

**Neutron scattering experiments.** Single crystal $Ce_2Zr_2O_7$ used for our experiments was described in detail in an earlier work[17]. Our polarized neutron scattering experiments were performed on the ThALES - Three Axis Low Energy Spectrometer at Institut Laue Langevin (ILL), Grenoble, France. One piece of single crystal (~ 2 grams) was mounted on a copper sample holder (Extended Fig. 1a). The crystal was pre-aligned using a lab X-ray Laue machine (Extended Fig. 1b), then aligned precisely using the OrientExpress neutron Laue station at ILL. The mosaics of the (0,0,4) and (1,1,1) nuclear Bragg peaks were $1.87 \pm 0.06$ and $2.37 \pm 0.24$ degrees FHWM, respectively. The momentum transfer $\boldsymbol{Q}$ in 3D reciprocal space in Å$^{-1}$ was defined as $\boldsymbol{Q} = h\boldsymbol{a}^* + k\boldsymbol{b}^* + l\boldsymbol{c}^*$, where $h$, $k$, and $l$ are Miller indices and $\boldsymbol{a}^* = 2\pi(\boldsymbol{b}\times\boldsymbol{c})/[\boldsymbol{a}\cdot(\boldsymbol{b}\times\boldsymbol{c})]$, $\boldsymbol{b}^* = 2\pi(\boldsymbol{c}\times\boldsymbol{a})/[\boldsymbol{a}\cdot(\boldsymbol{b}\times\boldsymbol{c})]$, $\boldsymbol{c}^* = 2\pi(\boldsymbol{a}\times\boldsymbol{b})/[\boldsymbol{a}\cdot(\boldsymbol{b}\times\boldsymbol{c})]$ with $\boldsymbol{a} = a_0\hat{\boldsymbol{x}}$, $\boldsymbol{b} = a_0\hat{\boldsymbol{y}}$, $\boldsymbol{c} = a_0\hat{\boldsymbol{z}}$ ($a_0 = 10.70$ Å at room temperature) and the $Fd\bar{3}m$ space group (Fig. 1a). The single crystal is aligned in the $(h,h,0) \times (0,0,l)$ scattering plane and mounted inside a dilution refrigerator kept at $T = 50$ mK for the entire experiment (Fig. 1b). The final neutron energy was fixed at $E_f = 3.23$ or 2.51 meV as specified in the figure caption. Both the monochromator and analyzer are horizontally and vertically focusing Heusler (1,1,1) crystals to produce and detect polarized neutrons. A flipping ratio measured on the (1,1,1) nuclear peak was $R = \frac{NSF_N}{SF_N} \approx \frac{20040}{650} \approx 30.8$. Scan at each point/energy takes ~35 seconds for NSF channels and ~175 seconds for SF channels. Due to the low incident neutron energies, no neutron filter was used.

The $x, y, z$ neutron polarization directions are defined as parallel to $\boldsymbol{Q}$, perpendicular to $\boldsymbol{Q}$ but in the scattering plane, perpendicular to $\boldsymbol{Q}$ and the scattering plane, respectively (Fig. 1c). The corresponding NSF and SF neutron scattering cross sections are summarized in Eq. (3). For polarized neutron scattering (elastic or inelastic) experiments with only vertical ($z$-axis) neutron polarization, the SF and NSF neutron scattering cross sections are $\sigma_z^{SF} = \frac{R}{R+1}M_y + \frac{1}{R+1}(M_z + N) + \frac{1}{R+1}\left[\frac{(2R+1)}{3}NSI + (R+1)B\right]$ and $\sigma_z^{NSF} = \frac{1}{R+1}M_y + \frac{R}{R+1}(M_z + N) + \frac{1}{R+1}\left[\frac{(R+2)}{3}NSI + (R+1)B\right]$, respectively.

Using full neutron polarization analysis by measuring $\sigma^{SF}_{x,y,z}(\mathbf{Q},E)$ and $\sigma^{NSF}_x(\mathbf{Q},E)$ at accessible of $\mathbf{Q}$ and $E$, one can determine $M_y$ and $M_z$, without the need to change temperature. In particular, $M_y$ and $M_z$ can be determined solely from $\sigma^{SF}_{x,y,z}(\mathbf{Q},E)$ using the first three rows of Eq. (3), which is exact for a paramagnet like Ce$_2$Zr$_2$O$_7$. We have

$$\sigma^{SF}_x = \frac{R}{R+1} M_y + \frac{R}{R+1} M_z + C_1 \quad (4)$$

$$\sigma^{SF}_y = \frac{1}{R+1} M_y + \frac{R}{R+1} M_z + C_1 \quad (5)$$

$$\sigma^{SF}_z = \frac{R}{R+1} M_y + \frac{1}{R+1} M_z + C_1 \quad (6)$$

where $C_1 = \frac{1}{R+1} N + \frac{(2R+1)/3}{R+1} NSI + B$

(4) – (5) gives $M_z = \frac{R+1}{R-1} (\sigma^{SF}_x - \sigma^{SF}_z) \quad (7)$

(4) – (6) gives $M_y = \frac{R+1}{R-1} (\sigma^{SF}_x - \sigma^{SF}_y) \quad (8)$

(7) + (8) gives $M_y + M_z = \frac{R+1}{R-1} (2\sigma^{SF}_x - \sigma^{SF}_y - \sigma^{SF}_z) \quad (9)$

(7) – (8) gives $M_z - M_y = \frac{R+1}{R-1} (\sigma^{SF}_y - \sigma^{SF}_z) \quad (10)$

The flipping ratio, $R \sim 30$, is a typical value from experience on Thales. In principle, there is an uncertainty of $R$ of a few percent, but in our case that would be a very small effect. A 10 % uncertainty of $R$ (from 30 to 27) would give a 0.6 % uncertainty of $\frac{R+1}{R-1}$ (from 1.069 to 1.077) $M_y$ or $M_z$. A 50 % uncertainty of $R$ (from 30 to 15) would only give a 7 % uncertainty of $\frac{R+1}{R-1}$ (from 1.069 to 1.142) $M_y$ or $M_z$. For this reason, we can safely ignore uncertainties of $R$.

Using the last three rows of Eq (3), we can also extract $M_y$ and $M_z$ from $\sigma^{NSF}_{x,y,z}(\mathbf{Q},E)$.

$$\sigma^{NSF}_x = \frac{1}{R+1} M_y + \frac{1}{R+1} M_z + C_2 \quad (11)$$

$$\sigma^{NSF}_y = \frac{R}{R+1} M_y + \frac{1}{R+1} M_z + C_2 \quad (12)$$

$$\sigma^{NSF}_z = \frac{1}{R+1} M_y + \frac{R}{R+1} M_z + C_2 \quad (13)$$

where $C_2 = \frac{R}{R+1}N + \frac{(R+2)/3}{R+1}NSI + B$

(12) – (11) gives $M_y = \frac{R+1}{R-1}(\sigma_y^{NSF} - \sigma_x^{NSF})$ (14)

(13) – (11) gives $M_z = \frac{R+1}{R-1}(\sigma_z^{NSF} - \sigma_x^{NSF})$ (15)

(14) + (15) gives $M_y + M_z = \frac{R+1}{R-1}(-2\sigma_x^{NSF} + \sigma_y^{NSF} + \sigma_z^{NSF})$ (16)

(15) - (14) gives $M_z - M_y = \frac{R+1}{R-1}(\sigma_z^{NSF} - \sigma_y^{NSF})$ (17)

However, $\sigma_{x,y,z}^{NSF}(\boldsymbol{Q}, E)$ contains nuclear scattering ($N$), which is much larger than $NSI$, especially at the elastic line. Therefore, we will have much larger uncertainties in obtaining the values of $M_y$ and $M_z$ due to propagation of statistical errors (see Extended Table 2). In practice, one would not use $\sigma_{x,y,z}^{NSF}(\boldsymbol{Q}, E)$ to extract the magnetic signal. Also, our counting time for $\sigma_{x,y,z}^{NSF}(\boldsymbol{Q}, E)$ is only 1/5 of $\sigma_{x,y,z}^{SF}(\boldsymbol{Q}, E)$, resulting even larger errors when calculating $M_y$ and $M_z$.

To estimate the instrumental energy resolution, we fit energy scans of $\sigma_x^{NSF}$ at different $\boldsymbol{Q}$ as shown in Extended Fig. 2. For $E_f = 3.23$ meV, the instrumental energy resolution is about 0.065 meV at FHWM. For $E_f = 2.52$ meV, the energy resolution is 0.042 meV at FHWM. Extended Figure 3 compares the raw data of $\sigma_x^{NSF}(\boldsymbol{Q})$ with $\sigma_{x,y,z}^{SF}(\boldsymbol{Q})$ at the elastic position $E = 0 \pm 0.03$ meV. We see NSF scattering dominates the scattering signal at all probed $\boldsymbol{Q}$. Extended Figure 4 compares the raw data of $\sigma_x^{NSF}(\boldsymbol{Q})$ with $\sigma_{x,y,z}^{SF}(\boldsymbol{Q})$ at the inelastic position $E = 0.1 \pm 0.03$ meV. Here magnetic scattering in the SF channel is larger than nonmagnetic scattering in the NSF channel.

To compare our measurements discussed in Figs. 1-4 with energy integrated spin-spin correlation function $S^y(\mathbf{Q}) = \int M_y(\mathbf{Q}, E)dE$ and $S^z(\mathbf{Q}) = \int M_z(\mathbf{Q}, E)dE$ discussed before[20], we integrated $M_y(\mathbf{Q}, E)$ and $M_z(\mathbf{Q}, E)$ of Figs. 2 in energy at X and K points and compare the outcome with our theoretical model and Ref. [20] in Extended Table 1.

| This work | Experiments | Calculation | Ref. 21 |
|---|---|---|---|
| X: $\int M_z(E)dE$ | 1 | 1 | 1 |
| X: $\int M_y(E)dE$ | 0.977 | 1.000 | 1.5 |
| K: $\int M_z(E)dE$ | 0.992 | 0.983 | 2.125 |
| K: $\int M_y(E)dE$ | 0.892 | 0.824 | 0.875 |

**Extended Table 1.** Normalized intensity of the energy integrated from -0.1 to 0.3 meV at X and K points in the NSF and SF channel, compared with theoretical calculation.

From Extended Table 1, we see that our estimated $\int M_z(E)dE$ is about 50% smaller than that from $\sigma_z^{NSF}(\mathbf{Q})$ measurements[20]. From Eq. (3), we note that $\sigma_z^{NSF}(\mathbf{Q})$ is sensitive to nuclear scattering (including phonons, $N$) in addition to the usual background scattering ($B$). To understand the problem, we note all previous unpolarized[16,17] and vertical field polarized[20] neutron scattering experiments assume that the background and other nonmagnetic scattering is temperature independent between base and high temperature. Using our measured $\sigma_{x,y,z}^{SF}(\mathbf{Q}, E)$ and $\sigma_x^{NSF}(\mathbf{Q}, E)$, we can determine $N$, $NSI$, and $B$ at the elastic position at base temperature (50 mK) without the need for high-temperature data. At $\mathbf{Q} = (0, 0, 1)$ and $E = 0 \pm 0.03$ meV, we find $C_1 = \frac{1}{R+1}N + \frac{(2R+1)/3}{R+1}NSI + B = 326$, $C_2 = \frac{R}{R+1}N + \frac{(R+2)/3}{R+1}NSI + B = 603$, $M_y = 57.5$, and $M_z = 62$ (Extended Table 2). As

NSI scattering is typically pretty small, the nuclear scattering in the NSF channel is comparable to the background scattering. Therefore, $\sigma_z^{NSF}(\boldsymbol{Q})$ measurements contain considerable nuclear scattering that needs to be eliminated, and the intrinsic magnetic signal is about 10% of the nonmagnetic scattering.

In order to improve statistics and compare with theoretical calculations, we repeated 3 ~ 5 times the measurements of $\sigma_{x,y,z}^{SF}(\boldsymbol{Q}, E)$ and $\sigma_{x,y,z}^{NSF}(\boldsymbol{Q}, E)$ at $\boldsymbol{Q} = (0,0,1), (1,1,0), (0.5,0.5,0.5), (0,0,2), (0,0,3)$ (Figs. 1b and 3a) with $E = 0 \pm 0.03$ and $0.1 \pm 0.03$ meV. This allows an overall determination of $M_y$ and $M_z$. Extended Table 2 summarizes these results at the elastic position and our calculation of $M_y+M_z$, $M_y$, and $M_z$. Extended Table 3 compares the outcome with theoretical calculations. Extended Tables 4 and 5 summarize similar results at $E = 0.1 \pm 0.03$ meV. Overall, these results are consistent with those discussed in Figs. 2-4 of the main text.

| Q | $\sigma_x^{SF}$ | $\sigma_y^{SF}$ | $\sigma_z^{SF}$ | $\sigma_x^{NSF}$ | $\sigma_y^{NSF}$ | $\sigma_z^{NSF}$ | $M_y+M_z$ (SF) | $M_y$ (SF) | $M_z$ (SF) | $M_y+M_z$ (NSF) | $M_y$ (NSF) | $M_z$ (NSF) |
|---|---|---|---|---|---|---|---|---|---|---|---|---|
| (0, 0, 1) | 442.0 | 388.0 | 384.0 | 657.5 | 705.0 | 665.0 | 119.7 | 57.7 | 62.0 | 58.8 | 50.8 | 8.0 |
| Error | 10.7 | 9.8 | 10.0 | 28.6 | 29.7 | 28.8 | 21.3 | 15.0 | 15.2 | 61.9 | 44.1 | 44.3 |
| (0, 0, 2) | 545.3 | 500.8 | 472.7 | 4257.0 | 4255.0 | 4320.0 | 125.2 | 47.6 | 77.7 | 65.2 | -2.1 | 67.3 |
| Error | 9.5 | 9.1 | 8.9 | 72.9 | 72.9 | 73.6 | 19.2 | 13.7 | 13.5 | 156.2 | 110.2 | 110.7 |
| (0, 0, 3) | 375.8 | 324.3 | 325.8 | 768.8 | 855.0 | 805.0 | 108.5 | 55.1 | 53.4 | 130.8 | 92.1 | 38.7 |
| Error | 9.7 | 9.0 | 9.0 | 31.1 | 32.6 | 31.8 | 19.4 | 13.7 | 13.7 | 67.7 | 48.2 | 48.7 |
| (1/2, 1/2, 1/2) | 470.0 | 392.8 | 387.2 | 663.0 | 828.0 | 720.0 | 171.0 | 82.5 | 88.5 | 237.3 | 176.4 | 60.9 |
| Error | 10.8 | 9.9 | 9.8 | 28.8 | 32.2 | 30.0 | 21.4 | 15.2 | 15.1 | 64.1 | 46.2 | 47.0 |
| (1, 1, 0) | 393.5 | 369.5 | 329.5 | 582.0 | 593.0 | 563.0 | 94.1 | 25.7 | 68.4 | -8.6 | 11.8 | -20.3 |
| Error | 9.9 | 9.6 | 9.1 | 24.1 | 24.4 | 23.7 | 19.9 | 14.3 | 13.9 | 51.5 | 36.7 | 36.3 |

**Extended Table 2.** The average counts per 3 minutes of the NSF and SF channel in the $x, y, z$ polarized direction at different $Q$ points and the corresponding errors at the elastic line ($E = 0 \pm 0.03$ meV). The actual counting time of the NSF channels are 1/5 of the corresponding SF channels. The total magnetic signal in $M_y$ and $M_z$ channels were calculated using Eq. (7-9) (using SF channels only) and Eq. (14-16) (using NSF channels only). Errors are statistical errors of 1 standard deviation.

| $Q$ | $M_y$ (Exp.) | $M_z$ (Exp.) | $M_y$ (Cal.) | $M_z$ (Cal.) |
|---|---|---|---|---|
| (0, 0, 1) | 0.65 | 0.70 | 1.04 | 1.04 |
| (0, 0, 2) | 0.54 | 0.88 | 0.79 | 0.79 |
| (0, 0, 3) | 0.62 | 0.60 | 1.04 | 1.04 |
| (1/2, 1/2, 1/2) | 0.93 | 1.00 | 1.00 | 1.00 |
| (1, 1, 0) | 0.29 | 0.77 | 0.62 | 1.04 |

**Extended Table 3.** Normalized intensity of the magnetic signal in $M_y$ and $M_z$ channels at different $Q$ points, compared with the theoretical calculations at the elastic line ($E = 0 \pm 0.03$ meV).

| $Q$ | $\sigma_x^{SF}$ | $\sigma_y^{SF}$ | $\sigma_z^{SF}$ | $\sigma_x^{NSF}$ | $\sigma_y^{NSF}$ | $\sigma_z^{NSF}$ | $M_y + M_z$ (SF) | $M_y$ (SF) | $M_z$ (SF) | $M_y+M_z$ (NSF) | $M_y$ (NSF) | $M_z$ (NSF) |
|---|---|---|---|---|---|---|---|---|---|---|---|---|
| (0,0, 1) | 62.3 | 47.0 | 44.5 | 28.0 | 43.0 | 51.5 | 35.4 | 16.4 | 19.0 | 41.2 | 16.0 | 25.1 |
| Error | 3.2 | 2.8 | 2.7 | 6.9 | 8.5 | 9.2 | 6.2 | 4.4 | 4.4 | 17.0 | 11.7 | 12.3 |
| (0, 0, 2) | 63.0 | 44.2 | 44.4 | 55.0 | 101.5 | 100.0 | 40.0 | 20.1 | 19.9 | 97.8 | 49.7 | 48.1 |
| Error | 9.5 | 9.1 | 8.9 | 9.6 | 13.0 | 13.0 | 19.2 | 13.7 | 13.5 | 24.4 | 17.3 | 17.3 |
| (0, 0, 3) | 61.4 | 38.4 | 40.3 | 18.3 | 41.7 | 60.0 | 47.2 | 24.6 | 22.6 | 69.6 | 25.0 | 44.6 |
| Error | 3.0 | 2.3 | 2.4 | 5.6 | 8.3 | 10.1 | 5.5 | 3.9 | 3.9 | 16.3 | 10.7 | 12.3 |
| (1/2, 1/2, 1/2) | 76.0 | 51.0 | 49.0 | 26.7 | 48.0 | 46.5 | 55.6 | 26.7 | 28.9 | 43.9 | 22.8 | 21.2 |
| Error | 5.0 | 4.1 | 4.0 | 6.7 | 8.9 | 8.7 | 9.5 | 6.7 | 6.7 | 16.8 | 12.0 | 11.8 |
| (1, 1, 0) | 67.0 | 49.6 | 45.3 | 25.0 | 34.0 | 30.0 | 41.8 | 18.6 | 23.2 | 15.0 | 9.6 | 5.3 |
| Error | 2.6 | 2.2 | 2.1 | 6.5 | 7.6 | 7.2 | 4.9 | 3.5 | 3.5 | 14.9 | 10.7 | 10.3 |

**Extended Table 4.** The average counts per 3 minutes of the NSF and SF channels in the $x, y, z$ polarized direction at different $Q$ points and the corresponding errors at the inelastic line ($E = 0.1 \pm 0.03$ meV). The actual counting time of the NSF channels are 1/5 of the corresponding SF channels. The total magnetic signal and in $M_y$ and $M_z$ channels were calculated using Eq. (7-9) (using SF channels only) and Eq. (14-16) (using NSF channels only). Errors are statistical errors of 1 standard deviation.

| $Q$ | $M_y$ (Exp.) | $M_z$ (Exp.) | $M_y$ (Cal.) | $M_z$ (Cal.) |
|---|---|---|---|---|
| (0, 0, 1) | 0.57 | 0.66 | 1.01 | 1.01 |
| (0, 0, 2) | 0.70 | 0.69 | 1.13 | 1.13 |
| (0, 0, 3) | 0.85 | 0.78 | 1.01 | 1.01 |
| (1/2, 1/2, 1/2) | 0.93 | 1.00 | 1.00 | 1.00 |
| (1, 1, 0) | 0.64 | 0.80 | 0.91 | 1.01 |

**Extended Table 5.** Normalized intensity of the magnetic signals in $M_y$ and $M_z$ channels at different $Q$ points, compared with the theoretical calculations at the inelastic position ($E = 0.1 \pm 0.03$ meV).

**Problems with using high-temperature (~10 K) as background in unpolarized neutron scattering experiments.**

In our previous unpolarized neutron scattering experiment[17], we assumed that the magnetic scattering at 12 K is diffusive enough and would be $Q$ and $E$ independent, and can thus serve as the nonmagnetic background. However, the Bose population factor dictates that any bosonic excitations (acoustic phonons and other background scattering) within 1 meV would be populated at 12 K and suppressed at 50 mK (Extended Fig. 5a and 5b). This is especially important at elastic and quasielastic positions for $Ce_2Zr_2O_7$ where $N$ and $B$ have about 90% of the total scattering signal. From the comparison of the integrated intensity along the [0, 0, $l$] and [$h$, $h$, 0] directions at the elastic line at 35 mK and 12 K (Extended Fig. 5c, 5d), we clearly see that the intensity at 12 K is higher than that at 35 mK in most of the reciprocal space in the scattering plane, likely due to thermal induced quasielastic scattering at 12 K. Therefore, the magnetic excitations at energies near zero were overlooked by the incorrect over-subtraction in unpolarized neutron scattering experiments. Even for polarized neutron scattering experiments with only vertical neutron polarization[20], high-temperature measurements were used as background scattering and similar problems may occur. For CSI such as $Ho_2Ti_2O_7$ with large effective magnetic moment, the magnetic scattering is many times larger than that of $Ce_2Zr_2O_7$[12,16,17,43,44]. Therefore, polarized neutron scattering experiment with only vertical neutron polarization using $\sigma_z^{SF}(Q)$ and $\sigma_z^{NSF}(Q)$ can still accurately determine $M_y$ and $M_z$ due to the overwhelming magnetic signal compared with the magnitude of the $N$, $NSI$, and $B$ scattering[12]. This is incorrect for $Ce_2Zr_2O_7$ with effective $S = 1/2$ as shown in Figs. 2, 3, and Extended Table 2.

**Possible effect from chemical disorder**

As determined from an earlier work[17], there is a 4% anti-site disorder of Ce and Zr from X-ray diffraction. Though $Ce_2Zr_2O_7$ can be air-sensitive, our single crystal sample was stored in an Ar glove box before and after the experiments, and we find no evidence of oxidation. Our data for $Ce_2Zr_2O_7$ is consistent with inelastic neutron scattering data of powder sample $Ce_2Sn_2O_7$ for energies above 0.03 meV, which has less disorder[57]. Recent work on single crystals of $Ce_2Sn_2O_7$[23], which has about 3% B-site stuffing of $Ce^{4+}$ (comparable with the 4% in $Ce_2Zr_2O_7$), behave similarly as powder $Ce_2Sn_2O_7$ in terms of heat capacity and other properties. From this perspective, a few percent disorder in

Ce$_2$Zr$_2$O$_7$ should not induce the wave vector dependent scattering (Fig. 3e) and spin excitations should not be anisotropic in spin space (Fig. 3c).

**Theoretical calculation of the dynamical spin structure factor in $\pi$-D-QSI.**

Considering that we are interested in small momentum transfer, we assume that the octupolar magnetic form factor associated with the $\tau^x$ and $\tau^y$ moments can be neglected and that the magnetic form factor of $\tau^z$ is constant over the momentum transfers of interest. Then, the magnetic scattering along the local y- and z-axis have the generic form

$$M^{y(z)}(\mathbf{Q}, E) = \frac{C}{N_{u.c.}} \int dt \sum_{i,j} P_{ij}^{y(z)}(\mathbf{Q}) e^{i(Et + \mathbf{Q}\cdot(\mathbf{R_i}-\mathbf{R_j}))} \langle \tau_i^z(t) \tau_j^z(0) \rangle, \quad (18)$$

where $C$ is a global prefactor that depends on the parameters of the experiment under consideration (e.g., sample size, neutron flux), $R_i$ labels the position of site $i$, and $P_{ij}(\mathbf{Q})$ is a polarization factor that depends on the sublattices of sites $i$ and $j$, the momentum transfer $\mathbf{Q}$ and the direction of the magnetic scattering. Specifically, for $M_z$ and $M_y$ we have $P^z_{ij}(\mathbf{Q}) = (\hat{\mathbf{e}}_{i,z} \cdot \hat{\mathbf{z}}_{sc})(\hat{\mathbf{e}}_{j,z} \cdot \hat{\mathbf{z}}_{sc})$ and $P^y_{ij}(\mathbf{Q}) = \left(\hat{\mathbf{e}}_{i,z} \cdot \frac{\mathbf{Q}\times\hat{\mathbf{z}}_{sc}}{|\mathbf{Q}\times\hat{\mathbf{z}}_{sc}|}\right)\left(\hat{\mathbf{e}}_{j,z} \cdot \frac{\mathbf{Q}\times\hat{\mathbf{z}}_{sc}}{|\mathbf{Q}\times\hat{\mathbf{z}}_{sc}|}\right)$, respectively, where $\hat{\mathbf{z}}_{sc}$ is a unit vector perpendicular to the scattering plane and $\hat{\mathbf{e}}_{i,z}$ are the basis vectors along the local z-axis for the pseudospin frame at site $i$. Using the pseudospins in the XYZ model, the above expression can be rewritten as

$$M^{y(z)}(\mathbf{Q}, E) = \frac{C}{N_{u.c.}} \int dt \sum_{i,j} P_{ij}^{y(z)}(\mathbf{Q}) e^{i(Et + \mathbf{Q}\cdot(\mathbf{R_i}-\mathbf{R_j}))} \left(\cos^2\theta \langle S_i^z(t) S_j^z(0)\rangle + \sin^2\theta \langle S_i^x(t) S_j^x(0)\rangle\right). \quad (19)$$

With the slave-particle construction introduced in the main text, the dynamical averages for $\pi$-D-QSI become $\langle S_i^x S_j^x \rangle = \langle \mathcal{E}_{r_1,r_2} \mathcal{E}_{r_3,r_4} \rangle$, which gives the photon propagator. Here $r_1$,

$r_2$, $r_3$, and $r_4$ label the centers of tetrahedra such that $r_1$ and $r_3$ ($r_2$ and $r_4$) correspond to up (down) tetrahedra, and $i$ and $j$ sit at the middle of the bond $r_1 \to r_2$ and $r_3 \to r_4$, respectively. For the transverse correlation, we first rewrite the expression in terms of raising/lowering operators as $\langle S_i^x S_j^x \rangle = -\frac{1}{4}(\langle S_i^+ S_j^+ \rangle - \langle S_i^- S_j^+ \rangle - \langle S_i^+ S_j^- \rangle + \langle S_i^- S_j^- \rangle)$ before rewriting everything with spinon operators

$$\langle S_i^z S_j^z \rangle = -\frac{1}{16}\Big(\langle \Phi_{r_1}^\dagger e^{i\mathcal{A}_{r_1,r_2}} \Phi_{r_2} \Phi_{r_3}^\dagger e^{i\mathcal{A}_{r_3,r_4}} \Phi_{r_4} \rangle - \langle \Phi_{r_2}^\dagger e^{-i\mathcal{A}_{r_1,r_2}} \Phi_{r_1} \Phi_{r_3}^\dagger e^{i\mathcal{A}_{r_3,r_4}} \Phi_{r_4} \rangle -$$

$$\langle \Phi_{r_1}^\dagger e^{i\mathcal{A}_{r_1,r_2}} \Phi_{r_2} \Phi_{r_4}^\dagger e^{-i\mathcal{A}_{r_3,r_4}} \Phi_{r_3} \rangle + \langle \Phi_{r_2}^\dagger e^{-i\mathcal{A}_{r_1,r_2}} \Phi_{r_1} \Phi_{r_4}^\dagger e^{-i\mathcal{A}_{r_3,r_4}} \Phi_{r_3} \rangle\Big). \text{ (20)}$$

In the following subsections, we explain how to evaluate the electric field propagator using Gaussian quantum electrodynamics and the four spinons correlations using GMFT.

**Gauge mean-field theory**

To use GMFT, we first rewrite the XYZ model in terms of raising/lowering spin operators as

$$\mathcal{H} = \sum_{\langle i,j \rangle}\big[\mathcal{J}_{||} S_i^{||} S_j^{||} - \mathcal{J}_\pm (S_i^+ S_j^- + S_i^- S_j^+) + \mathcal{J}_{\pm\pm}(S_i^+ S_j^+ + S_i^- S_j^-)\big], \text{ (21)}$$

where, for π-D-QSI, $\mathcal{J}_{||} = \mathcal{J}_x$, $\mathcal{J}_\pm = -(\mathcal{J}_z + \mathcal{J}_y)/4$, and $\mathcal{J}_{\pm\pm} = (\mathcal{J}_z - \mathcal{J}_y)/4$. In GMFT, the initial spin-½ Hilbert space on the pyrochlore lattice is augmented to a new larger one where bosonic degrees of freedom are introduced on the parent (premedial) diamond lattice whose sites are centered on the initial tetrahedra. For this mapping to be exact, the discretized Gauss's law $Q_r = \eta_r \sum_{i \in r} S_i^{||}$ needs to be imposed on all tetrahedra, where the sum is over all four sites that are part of the tetrahedra and $\eta_r = +1(-1)$ if $r$ is an up(down) tetrahedron. After mapping the pseudospin component with the dominant coupling to the emergent electric field $S_i^{||} = \mathcal{E}_{r,r'}$, the above definition is interpreted as a lattice divergence (i.e., $Q = \nabla \cdot \mathcal{E}$) and corresponds to the number of spinons at this site. The boson raising and lowering operators can then be defined as $\Phi_r^\dagger = e^{i\phi_r}$ and $\Phi_r = e^{-i\phi_r}$, respectively, where $\phi_r$ is canonically conjugate to $Q_r$. These quantum rotors respect $|\Phi_r^\dagger \Phi_r| = 1$ by construction. Using the mapping $S_i^+ = \frac{1}{2}\Phi_r^\dagger e^{i\eta_r \mathcal{A}_{r,r'}} \Phi_{r'}$ and $S_i^{||} = \mathcal{E}_{r,r'}$ introduced in the main text, the Hamiltonian can be rewritten as an interacting quantum rotor model strongly coupled to a compact $U(1)$ gauge field.

To get a tractable model, we carry out three successive approximations. (1) The four boson interactions coming from the $U(1)$ symmetry-breaking term $\mathcal{J}_{\pm\pm}$ are decoupled as $\Phi_i^\dagger \Phi_i^\dagger \Phi_j \Phi_k \rightarrow \Phi_i^\dagger \Phi_i^\dagger \chi_{j,k} + \Phi_j \Phi_k (\chi_{i,i}^0)^* + 2\Phi_i^\dagger \Phi_j \xi_{i,k} + 2\Phi_i^\dagger \Phi_k \xi_{i,j}$, where $\chi$, $\chi^0$, and $\xi$ are mean-field parameters representing inter-site pairing, on-site pairing, and inter-sublattice hopping, respectively. (2) The bosonic matter and dynamical gauge field sectors are decoupled by fixing the gauge field to a constant background $\mathcal{A} \rightarrow \overline{\mathcal{A}}$, where we pick a gauge configuration such that $\nabla \times \overline{\mathcal{A}} = \pi$ in the π-flux QSI phase and $\nabla \times \overline{\mathcal{A}} = 0$ in the 0-flux phase. (3) We relax the constraint $|\Phi_r^\dagger \Phi_r| = 1$ to the average one $\langle \Phi_r^\dagger \Phi_r \rangle = \kappa$ by performing a large-$N$ approximation. $\langle A \rangle$ denotes a thermal average, and the constraint is imposed by tuning a Lagrange multiplier $\lambda$. $\kappa = 2$ is chosen since such a constraint recovers the correct spinon dispersion in the Ising limit and agrees with quantum Monte-Carlo (QMC) results for the position of the phase transition from the 0-flux QSI to an ordered state[29,30]. Such a choice also agrees with QMC and exact diagonalization (ED) for the position of the lower and upper edges of the two-spinon continuum for 0-flux and π-flux QSI[30]. It should be emphasized that despite the apparent severity of such approximations, GMFT has been extensively benchmarked and shown to give great qualitative agreement with state-of-the-art QMC, ED, and pseudofermions functional renormalization group[30]. After such approximations, we have a non-interacting quadratic Hamiltonian that can be diagonalized exactly. When solving the self-consistency conditions in the relevant parameter regimes, it is found that the mean-field parameters all vanish in the deconfined phase ($\chi = 0$, $\xi = 0$, and $\chi^0 = 0$). GMFT is thus insensitive to $\mathcal{J}_{\pm\pm}$ for the parameter regime of interest. Consequently, only $\mathcal{J}_{\parallel}$ and $\mathcal{J}_{\pm}$ are fitted when comparing with experiments. From this quadratic Hamiltonian, the dynamical spin structure factor can be evaluated as in Ref. [30].

**Gaussian quantum electrodynamics**

The low-energy physics of the spin ice manifold can be described by a compact U(1) gauge theory

$$\mathcal{H}_{\mathcal{U}(1)} = \frac{\mathcal{U}}{2}\sum_{\langle r,r'\rangle} \mathcal{E}_{r,r'}^2 - \mathcal{K} \sum_h \cos(\nabla \times \mathcal{A})_h, \quad (22)$$

where the second sum is over hexagonal plaquettes. In its deconfined phase (i.e., for QSI), the low-energy physics of a compact U(1) gauge theory can be approximated by

$$\mathcal{H}_{\mathcal{U}(1)} \approx \frac{\mathcal{U}}{2}\sum_{\langle r,r'\rangle} \mathcal{E}_{r,r'}^2 + \frac{\mathcal{K}}{2}\sum_h (\nabla \times \mathcal{A})_h^2. \quad (23)$$

This quadratic Hamiltonian can be diagonalized to compute the dynamical spin structure factor at finite temperature due to photons, as explained in Ref. [15]. This procedure yields

$$\int dt \sum_{i,j} e^{i(Et + \mathbf{Q}\cdot(\mathbf{R_i}-\mathbf{R_j}))} \langle \mathcal{E}_{r_1,r_2}(t) \mathcal{E}_{r_3,r_4}(0) \rangle$$

$$= \frac{\zeta^2 \mathcal{K}}{2E(\mathbf{Q})} \sum_{\mu\nu} \sum_\lambda \sin(\mathbf{Q}\cdot h_{\mu\nu}) \sin(\mathbf{Q}\cdot h_{\nu\lambda}) \Big[ n_B(E(\mathbf{Q}))\delta(E + E(\mathbf{Q})) + \Big(1 + n_B(E(\mathbf{Q}))\Big) \delta(E - E(\mathbf{Q}))\Big], \quad (24)$$

where $\mu, \nu, \lambda \in \{0,1,2,3\}$ label the four pyrochlore sublattices, $h_{\mu\nu} = a_0(b_\mu \times b_\nu)/(\sqrt{8}|b_\mu \times b_\nu|)$ where $b_\mu$ are the vectors connecting an up tetrahedron to its four nearest-neighbor down tetrahedra, $n_B$ is the Bose-Einstein distribution and $E(\mathbf{Q})$ is the photon dispersion. $\zeta$ is a dimensionless parameter that is meant to take into account any renormalization of the electric field when integrating out high-energy degrees of freedom to derive the above effective field theory. We set $\zeta = 1$. Any other choice would affect the relative intensity of the photon and spinon contributions and the fitted θ value.

The parameters $\mathcal{U}$ and $\mathcal{K}$ are related to the speed of the emergent light by $c_{QSI} = \sqrt{\mathcal{U}\mathcal{K}} a_0 \hbar^{-1}$. The speed of light fully determines the photon dispersion in this description, but the intensity of the dynamical spin structure factor depends on the specific ratio of $\mathcal{U}$ and $\mathcal{K}$ [15]. Both $\mathcal{U}$ and $\mathcal{K}$ thus need to be fixed. In the perturbative regime close to the Ising point, one can relate $\mathcal{K}$ to the perturbative ring exchange term obtained by going to the third order in perturbation theory (i.e., $\mathcal{K} \sim \mathcal{J}_\pm^3/\mathcal{J}_\parallel^2$) and fix $\mathcal{U}$ by using previous quantum Monte Carlo results[15]. However, since we are far from this perturbative regime, we make no such assumption regarding possible connections between parameters of the effective theory and the microscopic couplings of the initial spin Hamiltonian. Instead, we fix $c_{QSI}$ and then use the results of Ref. [49], where the effective ring exchange model of Eq. (23) was studied using large-scale exact diagonalization simulations. This investigation reports that $\hbar c_{QSI}/a_0 = 0.51(6) \mathcal{K}/2$. We thus use these results to fix $\mathcal{U}$ and $\mathcal{K}$ given $c_{QSI}$.

**Exploration of alternative explanations for the quasielastic signal.**

1. Thermal spinons: Within GMFT, spinons can also produce a quasielastic signal at finite temperatures. Indeed, at finite temperatures, instead of exciting two spinons, a neutron can be scattered by deexciting a thermally excited spinon and exciting another one from the vacuum for a net approximately null energy transfer. However, we find that this quasielastic signal from thermal spinons is systematically much smaller than the inelastic contribution for the experimental temperature of $T = 50$ mK as seen in Fig. 1d. It thus appears implausible that thermal spinons could yield the dominant quasielastic signal we report.

2. A broader spinon continuum: A natural question is whether one can fit the whole signal by simply lowering the spinon gap without invoking the emergent photon. It should first be noted that in a deconfined phase (i.e., QSI), the bosonic spinons need to have a finite gap $\Delta_{spinon}$. A gapless bosonic spinon dispersion would lead to spinon condensation and thus indicate a transition to a magnetically long-range ordered state. Inelastic neutron scattering probes the two-spinon continuum. As such, it is only non-zero for energies above the lower edge of the two-spinon continuum $2\Delta_{spinon}$ (in the non-interacting limit). Suppose only the spinons contribute to the dynamical spin structure factor. In that case, we then only expect an inelastic signal where quasielastic contribution would come from leaking of the inelastic contribution due to finite energy resolution (and the small contribution from thermal spinons mentioned above). Provided the energy step size in the signal measured is sufficiently small compared to the gap, the signal's maximum should then necessarily be for a non-zero energy transfer. This already seems at odds with our measurements, where we systematically see the maximum of the total magnetic signal

$M_z + M_y$ at (or very close to) the elastic line for the three momentum transfers reported in Fig. 2.

Before exploring further, it should also be emphasized that we believe GMFT to yield reasonable estimates for the position of the two-spinon continuum. Indeed, it gives good agreement for the position of the continuum with the QMC results of Ref. [46] for 0-QSI and the 32-site ED results of Ref.[28] for π-QSI, as seen in Extended Figure 6. The microscopic exchange couplings of Ce$_2$Zr$_2$O$_7$ have already been estimated through a detailed examination of thermodynamics and neutron scattering measurements[20,27]. The GMFT prediction for the spinon contribution to the dynamical spin structure factor using these parameters yields a first peak at approximately 0.05 meV, the second one around 0.1 meV[30]. This is in agreement with our measurements and interpretation of the energy scans in Fig. 2. Considering both the agreement for the predicted spinon gap of GMFT, ED, and QMC and the parameter sets that have been put forward through careful analysis of thermodynamic measurements, a spinon peak at much lower energies is unlikely. A much lower spinon peak would probably yield predictions incompatible with the system's thermodynamic behavior.

Nonetheless, we still tried to see if the results could be fitted by a spinon signal with a smaller gap without invoking any supplemental quasielastic contribution. To this end, we can reduce the position of the spinon peak by increasing $\mathcal{J}_\pm/\mathcal{J}_\parallel$ and decreasing $\mathcal{J}_\parallel$ (see Extended Figure 6). In particular, using $\mathcal{J}_\parallel = 0.06$ meV and $\mathcal{J}_\pm/\mathcal{J}_\parallel = -0.35$ produces a signal where the first spinon peak is around 0.03 meV as shown in Extended Figure 7. Such a spinon signal systematically fails to capture the large contribution at the elastic line. This is especially apparent when looking at the residual, where we clearly see a large signal missed at the elastic line. This residual can be accounted for using a Gaussian function centered around $E \approx 0$ meV with a width comparable to the experimental resolution. This is highly suggestive that there is an additional mode near zero energy regardless of any other consideration.

In short, it does not seem likely that our measurements can be accounted for using a spinon signal with a much smaller gap because this would likely be incompatible with the previously established thermodynamic behavior of $Ce_2Zr_2O_7$. Ignoring these constraints, we further find that a spinon-only fit is unable to account for the dominant quasielastic signal we report.

3. Magnetic disorder: Considering the importance disorder can have on the low-energy physics of highly frustrated magnets, one could be concerned that the signal might be due to disorder. However, the quasielastic signal differs significantly from what is expected of conventional magnetic disorder. First, the magnetic scattering is expected to have very weak wave vector dependence for impurity-induced spin glass state. For example, in classical spin glass such as $Cu_{1-x}Mn_x$ with x=0.0165 and 0.033, AC susceptibility shows clear hysteresis, and the imaginary part of dynamic susceptibility is weakly $\boldsymbol{Q}$-dependent and has a broad peak near 2 meV[65]. This is clearly different from $Ce_2Zr_2O_7$ where there is no frequency-dependent AC susceptibility, and magnetic scattering is momentum-dependent. This is especially clear by looking at $M_y$ in Fig. 3 where we see strong scattering along the First Brillouin zone boundary and a slow decrease of the intensity for larger values of $h$ in the $[h,h,0]$ direction. Next, for magnetic disorder such as in spin glasses where spin directions are randomly frozen in space, one expects spin excitations to be isotropic in spin space (i.e., $M_y \approx M_z$ for all $\mathbf{Q}$). In contrast, the quasielastic magnetic scattering from $Ce_2Zr_2O_7$ is isotropic at certain points, but anisotropic with $M_z > M_y$ at others. This is clear from Figs. 3a-3c. The spin space anisotropy is also momentum dependent, thus giving further support to our previous point about the momentum dependence of the quasielastic signal. For these reasons, it is quite challenging to reconcile our measurements with magnetic disorder. In contrast, our modeling in terms of emergent quantum electrodynamics captures the observed momentum and polarization dependence.

**Fits and comparison between theoretical predictions and experiments**

When comparing experimental and theoretical results, the theoretical prediction is convolved with a Gaussian distribution to consider finite experimental resolution. For example, when computing the theoretical predictions for the momentum scans at the elastic position $E = 0$ meV and $E = 0.1$ meV, the theoretical predictions are convolved with a Gaussian with the same full width at half maximum as the experimental resolution. We also fix the global prefactor $C$ introduced in Eq. (19) to the one that yields the best agreement with experiments. $C$ is chosen consistently between the theoretical calculations reported in Figs. 2, 3, and 4.

We impose a thermodynamics constraint on the speed of emergent light. Ref. [20] reported that the heat capacity down to $T_{\text{exp.}} = 50$ mK cannot be smoothly extrapolated with a cubic function. Using the photon dispersion of Gaussian quantum electrodynamics, the photon contribution to the heat capacity scales as $T^3$ up to about $k_B T = \hbar c/(2a_0)$. This naturally implies that $\hbar c/a_0 < 2k_B(50 \text{ mK}) \approx 0.009$ meV. We take this as an upper bound on the speed of the emergent photons.

To obtain the optimal parameter sets, we minimize the goodness of fit measure $\chi^2 = \sum_E (I_{E,Q}^{\text{Theory}} - I_{E,Q}^{\text{Experiment}})^2 / \Delta I_{E,Q}^{\text{Experiment}}$ where $I_{E,Q}^{\text{Experiment}}$ is the intensity of the signal (i.e., $M_z + M_y$ and $M_z - M_y$) measured experimentally at energy $E$ for the energy scans at the three different momentum points Q presented in Fig. 2 and $\Delta I_{E,Q}^{\text{Experiment}}$ is the associated uncertainty. For this fit, we do not fix the width of that Gaussian we convolve the results with, but use it as a free parameter to be optimized. We allow the FWHM of this broadening function to be within 50 percent of the experimental resolution. Such a fitting procedure yields $\mathcal{J}_x = 0.076$ meV and $\mathcal{J}_\pm = 0.021$ meV. These parameters are approximately 1.2 times the ones in Ref. [20] (i.e., the ratio $\mathcal{J}_\pm/\mathcal{J}_x$ is the same). We find that the goodness of fit is very shallow as a function of $c_{\text{QSI}}$. An extended range of values of $c_{\text{QSI}}$ and $\theta$ yields very reasonable and similar fits. In cases where the photon energy for small $c_{\text{QSI}}$ is much smaller than the experimental resolution, the photon signal essentially yields a Gaussian centered at the elastic position after the convolution procedure outlined above. Therefore, $c_{\text{QSI}}$ and $\theta$ cannot be uniquely determined with the energy scans of Fig. 2. We find acceptable agreement for $\hbar c_{\text{QSI}}/a_0 \in [k_B T_{exp.}/10, 0.65 k_B T_{exp.}] = [0.0004, 0.0028]$ meV. The lower bound on the speed of the emergent light is introduced since, for lower values, the dynamics of the 2-in-2-out manifold should be completely classical. For every speed of light in this interval, we determine an optimal $\theta$ by requiring that the ratio of the spinon and photon signals remains consistent. These optimal values of $\theta$ are in the range $[0.05\pi, 0.12\pi]$.

To determine the optimal speed of light within the above parameter range, we fit the elastic scans of $M_y$ and $M_z$ presented in Fig. 3 using a similarly defined goodness of fit $\chi^2 = \sum_Q (I_{E=0,Q}^{\text{Theory}} - I_{E=0,Q}^{\text{Experiment}})^2 / \Delta I_{E=0,Q}^{\text{Experiment}}$ where $I_{E=0,Q}^{\text{Experiment}}$ is now the intensity of the elastic signal (i.e., $M_y$ and $M_z$ at $E = 0 \pm 0.03$ meV) measured experimentally at the momentum transfer $Q$. The optimal speed of light saturates the upper bound defined from fitting the energy scans of Fig. 2 such that $\hbar c_{\text{QSI}}/a_0 = 0.65 \, k_B T_{exp.} = 0.0028$ meV. If we do not impose the above upper bound on $c_{\text{QSI}}$, we find that the goodness of fit for the elastic scan is monotonically decreasing with the speed of light up to about $\hbar c_{\text{QSI}}/a_0 \approx k_B T_{exp.}$. In summary, the optimal speed of light consistent with both the energy scans presented in Fig. 2 and the elastic momentum scans of Fig. 3 is $\hbar c_{\text{QSI}}/a_0 = 0.65 \, k_B T_{exp}$. The corresponding value of the angle is $\theta = 0.12\pi$.

Of all the parameters we estimate, $\theta$ is the most uncertain since it relies on a quantitive comparison between the dynamical spin structure factor obtained by GMFT and Gaussian quantum electrodynamics. It remains unclear to what extent this quantitative comparison between the two is accurate. The main point of the theoretical fit is not to offer a quantitative estimate of $\theta$, but rather to reproduce the experimental results using a small value of $\theta$ as well as $\mathcal{J}_x$ and $\mathcal{J}_\pm$ that are reasonably consistent with previous work[20,27]. Furthermore, it should be mentioned that previous QMC investigations on 0-flux QSI at small (i.e., $T < 12\,\mathcal{J}_\pm^3/\mathcal{J}_\|^2$) and intermediate temperatures (i.e., $12\mathcal{J}_\pm^3/\mathcal{J}_\|^2 < T < \mathcal{J}_\|/2$) found that the photon contribution to the dynamical spin structure factor is about one to four orders of magnitudes larger than the inelastic spinon contribution[46]. Such intensity ratios between the spinon and the photon contribution are consistent with our modeling. The estimated $\theta$ should thus be of the right magnitude.

**Supplementary Information** is linked to the online version of the paper.

**Acknowledgements**

We thank Han Yan and Andriy Nevidomskyy for helpful discussions. The neutron scattering work at Rice is supported US DOE BES DE-SC0012311 (P.D.). The single crystal growth work at Rice is supported by the Robert A. Welch Foundation under Grant No. C-1839 (P.D.). Crystal growth by B.G. and S.W.C. at Rutgers was supported by the visitor program at the center for Quantum Materials Synthesis (cQMS), funded by the Gordon and Betty Moore Foundation's EPiQS initiative through grant GBMF6402, and by Rutgers University.  F.D. and Y.B.K. are supported by the Natural Science and Engineering Research Council of Canada and the Center for Quantum Materials at the University of Toronto. D.W.T. acknowledges funding from the European Union's Horizon 2020 research and innovation programme under the Marie Skłodowska-Curie grant agreement No 884104 (PSI-FELLOW-III-3i). Neutron data (https://doi.ill.fr/10.5291/ILL-DATA.4-05-851) were obtained using OrientExpress and Thales instruments at ILL with support from proposal 4-05-851.

**Author Contributions**

P.D. and B.G. conceived the project. B.G., F.D. and D.W.T. made equal contributions to the project. B.G. and S.-W.C. prepared the samples. Neutron scattering experiments were carried out and analyzed by D.W.T., P.S., A.H., Y.S., B.G., and P.D.. Theoretical analysis was supervised by Y.B.K. and performed by F.D. and Y.B.K.. The entire project was supervised by P.D. The manuscript is written by P.D., Y.B.K., B.G., and F.D.. All authors made comments. The authors declare no competing financial interests.

Correspondence and requests for materials should be addressed to Y.B.K. (yongbaek.kim@utoronto.ca) or P.D. (pdai@rice.edu).

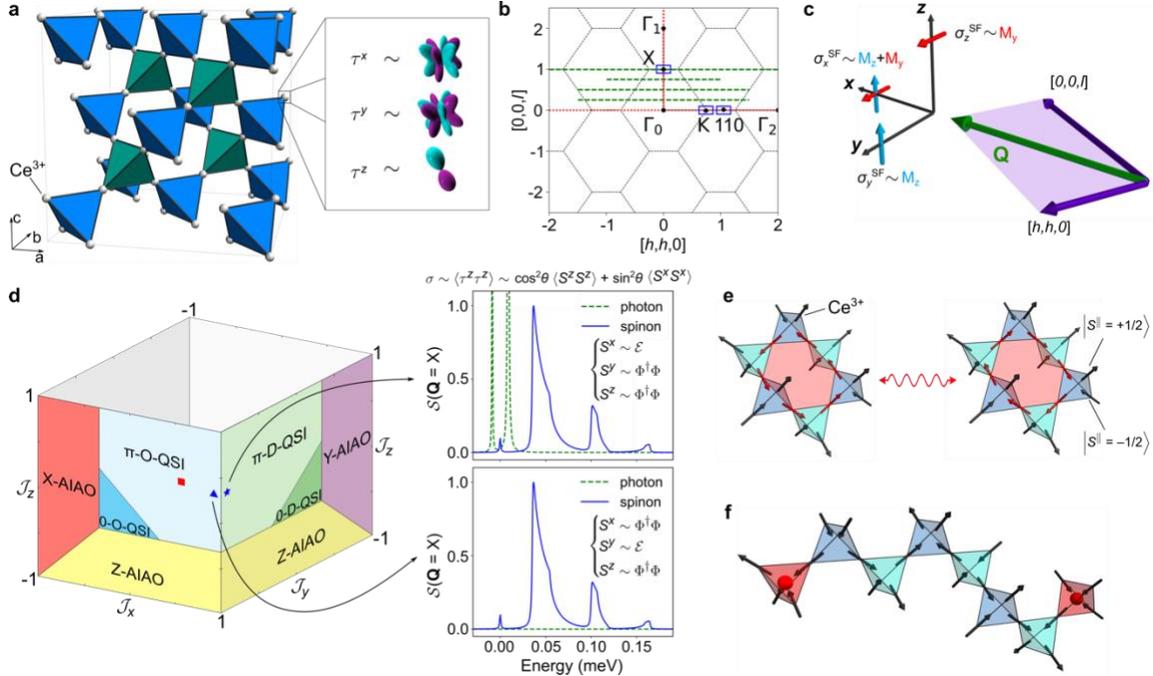

**Figure 1 | Summary of crystal structure, Brillouin zone, phase diagram, expected experimental signatures, and excitations of $Ce_2Zr_2O_7$. a,** Network of corner-sharing tetrahedra formed by the magnetic $Ce^{3+}$ ions in $Ce_2Zr_2O_7$. The low-energy Kramers doublet of the $Ce^{3+}$ ion can be modeled by the pseudospins-½ $\vec{\tau} = (\tau^x, \tau^y, \tau^z)$ that have octupolar and dipolar magnetic charge densities. **b,** Schematic diagram of the $(h, h, l)$ plane. The dashed black lines mark the zone boundary, the blue boxes the momentum transfer of the energy scans reported in Fig. 2, the green dashed lines the momentums along which the elastic scans presented in Fig. 3 were measured, and the red dotted lines the momentum cuts along which both the elastic and inelastic scans presented in Figs. 3 and 4 were measured. **c,** Scattering geometry of polarized neutron scattering experiment in the $(h, h, l)$ plane. Incident neutron beams are polarized along the $x$, $y$, and $z$ directions, corresponding to directions along the momentum transfer $\mathbf{Q}$, perpendicular to $\mathbf{Q}$ but in the $(h, h, l)$ plane, and perpendicular to the scattering plane, respectively. For this configuration, the SF cross section $\sigma_x^{SF}(\mathbf{Q}) \sim M_z + M_y$ where $M_y$ and $M_z$ are magnetic fluctuations along the local $y$ and $z$ directions, respectively, for large moment systems. Correspondingly, $\sigma_z^{SF}(\mathbf{Q}) \sim M_y$ and $\sigma_z^{NSF}(\mathbf{Q}) \sim M_z$. **d,** Phase diagram for the nearest-neighbor XYZ model[66] in the experimentally relevant quadrants for $Ce_2Zr_2O_7$. The labels X(Y)[Z]-AIAO represent all-in-all-out magnetic order along the local $x(y)[z]$ axes. The red square denotes the parameterization of Ref.[27], and the blue triangle and blue star are the ones of Ref.[55]. For the parameters in π-D-QSI, the photon can be observed in the neutron scattering cross section, but it cannot for parameters in the π-O-QSI phase. In both cases, the spinons produce a small quasielastic peak from thermal excitations and three inelastic peaks of decreasing intensities. **e,** Tunneling process between different spin ice configurations that form the coherent photon excitation. A vector pointing out of an up tetrahedron (in blue) is a representation of a positive pseudospin component for the one with the largest coupling $|S^\parallel = +1/2\rangle$ and a vector pointing out is $|S^\parallel = -1/2\rangle$. **f,** When flipping a spin, one

violates the ice rules and creates a spinon-antispinon pair that live at the center of the tetrahedra and are sources of the emergent electric field.

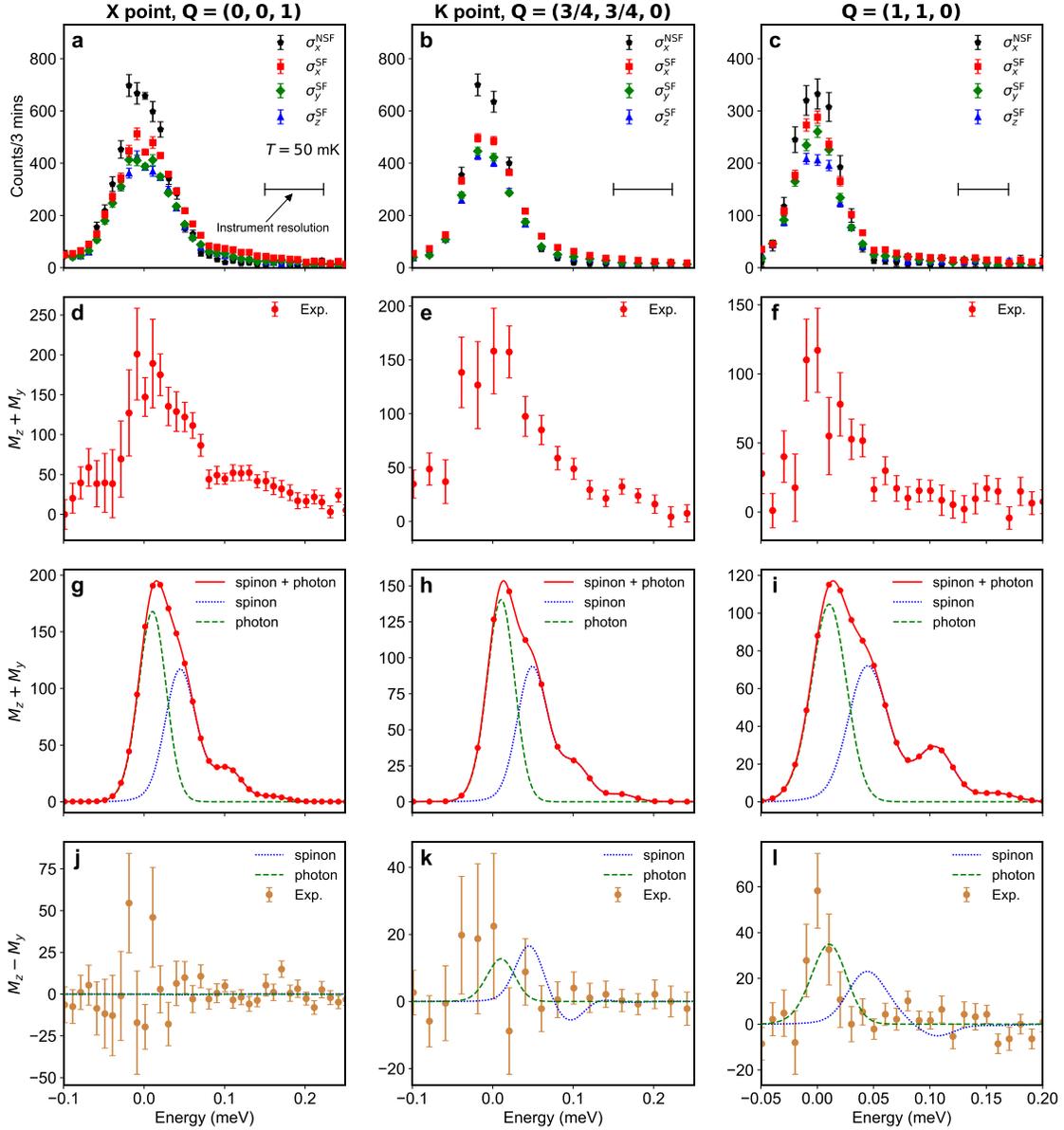

**Figure 2 | Energy scans of the polarized neutron scattering cross sections at different momentum positions for Ce$_2$Zr$_2$O$_7$. a-c**, Neutron scattering cross sections for different polarization as a function of energy at $Q = (0,0,1)$ (X point), $Q = (3/4,3/4,0)$ (K point), and $Q = (1,1,0)$. **d-f**, Total magnetic scattering $M_z + M_y$. **g-h**, Theoretical fit to $M_z + M_y$ using GMFT for the spinons and gaussian quantum electrodynamics for the photons. **j-l**, Polarization anisotropy $M_z - M_y$ and theoretical prediction for the spinons and photons contributions to it. The theoretical results were produced using $\mathcal{J}_x = 0.076$ meV, $(\mathcal{J}_y + \mathcal{J}_z)/4 = 0.021$ meV, $\theta = 0.12\pi$, and $\hbar c_{\text{QSI}}/a_0 = 0.65\, k_B T_{exp.}$. To incorporate finite experimental resolution, the results are broadened using a Gaussian with a full width at half

maximum (FWHM) of 0.04 meV, 0.04 meV, and 0.035 meV at the X, K and $Q = (1,1,0)$ point, respectively. The vertical dotted line in **d–i** are guides to the eye to denote the transitions from the quasielastic photon signal to the first spinon peak and between the three spinon peaks of π-flux QSI. The vertical error bars in **a–c** are statistical errors of 1 standard deviation. The horizontal bars in **a-c** are the instrumental energy resolutions in FHWM as determined from the energy width of $\sigma_x^{NSF}(E)$. Data in **a,b** are obtained with $E_f = 3.23$ meV, while in **c** are obtained with $E_f = 2.51$ meV. The vertical error bars in **d-f** and **j-l** are propagating errors using Eq. (3).

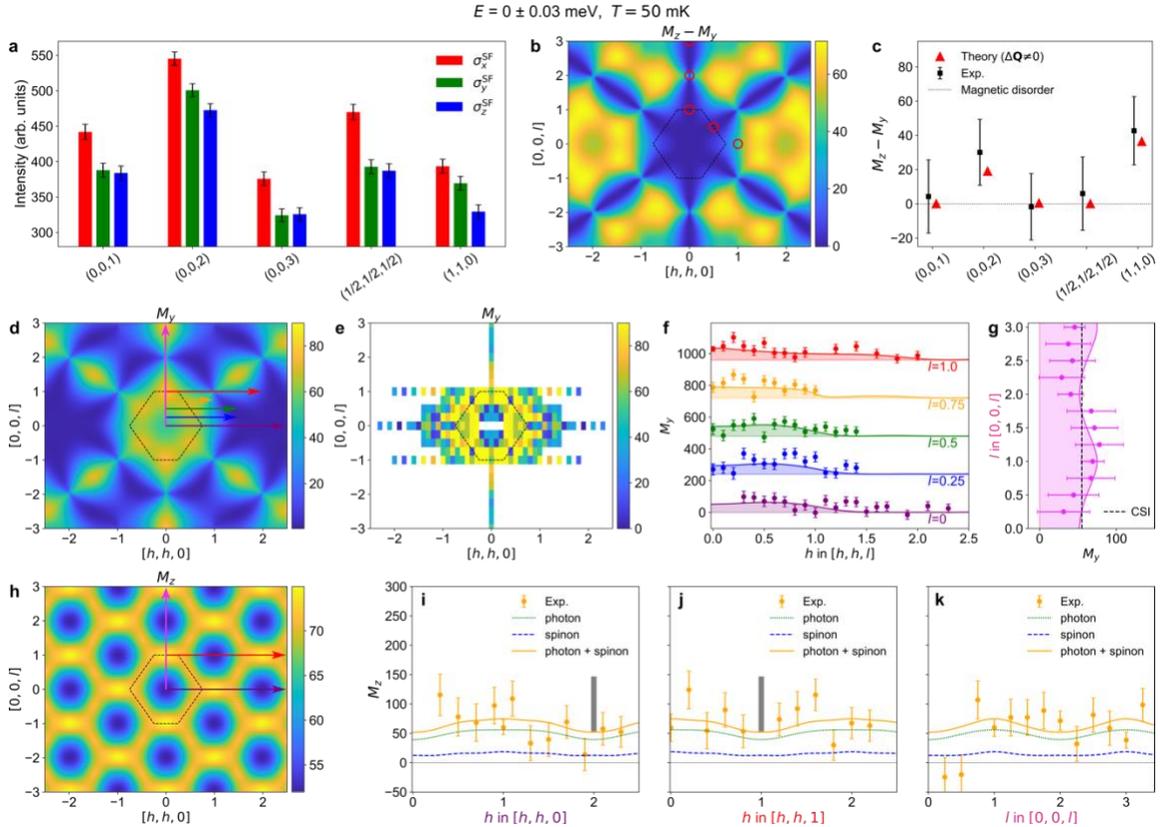

**Figure 3 | Wave vector and polarization dependence of the magnetic scattering at $E = 0 \pm 0.03$ meV for Ce$_2$Zr$_2$O$_7$. a,** $\sigma_{x,y,z}^{SF}(Q)$ at various momentum positions. **b,** Theoretical prediction for the polarization anisotropy $M_z - M_y$ from the total of spinons and photons at the elastic line $E = 0 \pm 0.03$ meV in the $[h, h, l]$ scattering plane using $\theta = 0.12\pi$, and $\hbar c_{QSI}/a_0 = 0.65\ k_B T_{exp.}$. The momentum points measured in **a** are highlighted in red. **c,** Comparison between theory with a small momentum uncertainty (i.e., $\Delta Q \neq 0$), experiments and expectations from conventional magnetic disorder for the polarization anisotropy. **d,e,** Theoretical predictions and measurements of $M_y$ in the $[h, h, l]$ scattering plane. The color bar in **e** is capped off to allow for better comparison with theory. **f,g,** Comparison between measurements and theoretical predictions for $M_y$ along the $[h, h, l]$ directions with $l=0, 0.25, 0.5, 0.75, 1$, and along the $[0, 0, l]$ direction. They are indicated by arrows in panel **d**. The results in **f** are shifted vertically for clarity. The results in **h** are

compared to CSI. **h,** Theoretical predictions of $M_z$ in the $[h, h, l]$ scattering plane. **i-k**, $M_z$ along the $[h, h, 0]$, $[h, h, 1]$ and $[0, 0, l]$ directions, and the theoretical calculation for the contributions from spinons, photons, and their total. These lines are indicated by arrows in panel **d**. Gray windows in panels **b**, **e**, and **f** indicate nuclear Bragg peaks at the (1, 1, 1) and (2, 2, 0) points. The vertical error bars in **a**, **c, e-g** and **i-k** are propagating errors using Eq. (3). Data in **a**, **c, e-g** and **i-k** are obtained with $E_f = 3.23$ meV. The vertical error bars in **a** are statistical errors of 1 standard deviation. They are propagating errors in **c**, **f**, **g**, **i-k** using Eq. (3).

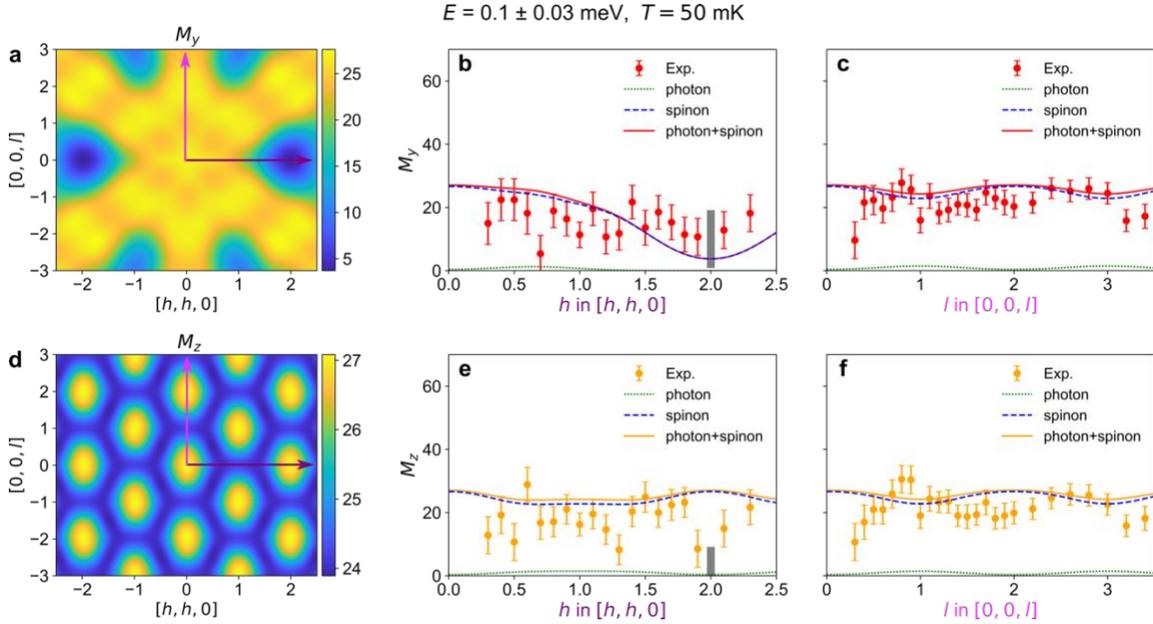

**Figure 4 | Wave vector and polarization dependence of magnetic scattering at $E = 0.1 \pm 0.03$ meV for Ce$_2$Zr$_2$O$_7$. a,d,** Theoretical predictions for $M_y$ and $M_z$ from the total of spinons and photons at $E = 0.1 \pm 0.03$ meV in the $[h, h, l]$ scattering plane using $\theta = 0.12\pi$, and $\hbar c_{QSI}/a_0 = 0.65\ k_B T_{exp.}$. **b,c,** $M_y$ along the $[h, h, 0]$ and $[0, 0, l]$ directions and the theoretical predictions for the contributions from spinons, photons, and their total. These lines are indicated by arrows in panel **a**. **e,f,** $M_z$ along the $[h, h, 0]$ and $[0, 0, l]$ directions and the theoretical predictions for the contributions from spinons, photons, and their total. These lines are indicated by arrows in panel **d**. Gray windows in panels **b** and **e** indicate the nuclear Bragg peak (2, 2, 0). The vertical error bars in **b**, **c**, and **e**, **f** are propagating errors using Eq. (3). Data in **b,c** and **e**, **f** are obtained with $E_f = 3.23$ meV.

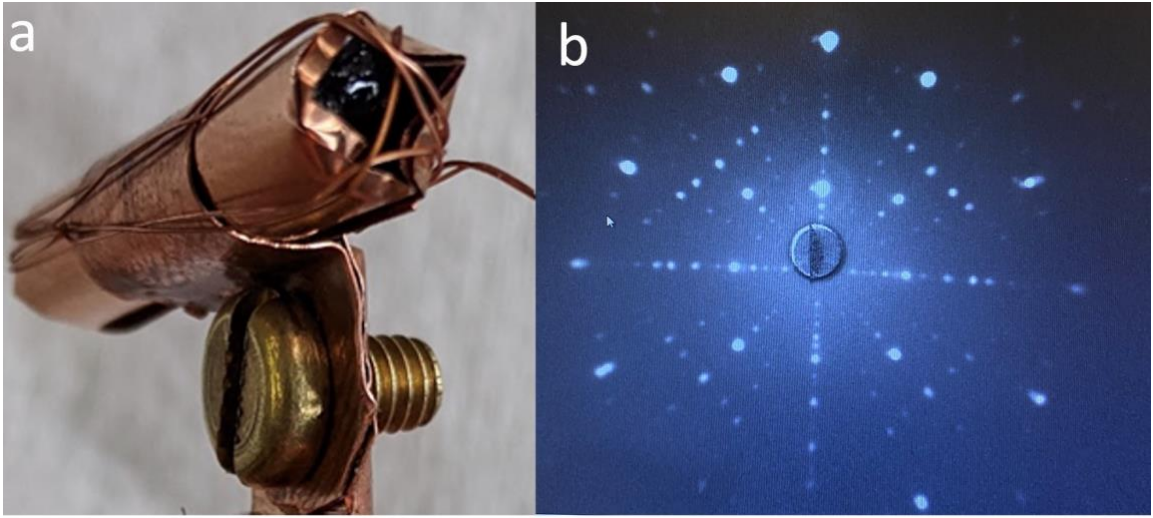

**Extended Figure 1**. **a**, One piece of single-crystalline $Ce_2Zr_2O_7$ was mounted on a copper holder. **b**, The X-ray Laue pattern in the [0, 0, 1] direction. The sample is mounted inside a dilution refrigerator maintained at $T = 50$ mK for the entire experiment. The sample is tied inside copper foils to ensure good thermos-conductivity at 50 mK.

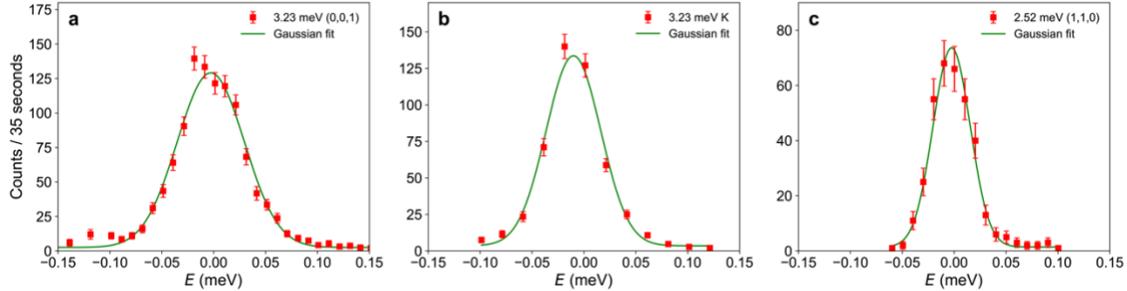

**Extended Figure 2**. **a-c**, The energy scan of $\sigma_x^{NSF}(Q, E)$ channel at $Q = (0,0,1), (3/4, 3/4, 0)$ and $(1,1,0)$ using $E_f = 3.23, 3.23,$ and $2.52$ meV, respectively. We use the Gaussian fit to determine the energy resolution to be 0.076 meV, 0.062 meV and 0.042 meV in FWHM, respectively. The actual instrument resolution may be better because $\sigma_x^{NSF}(Q, E)$ contains nuclear scattering from sample and environment, which can have phonon and other nonmagnetic excitations that broaden the intrinsic energy width, in addition to the instrumentation resolution limited NSI scattering. The vertical error bars in **a**–**c** are statistical errors of 1 standard deviation.

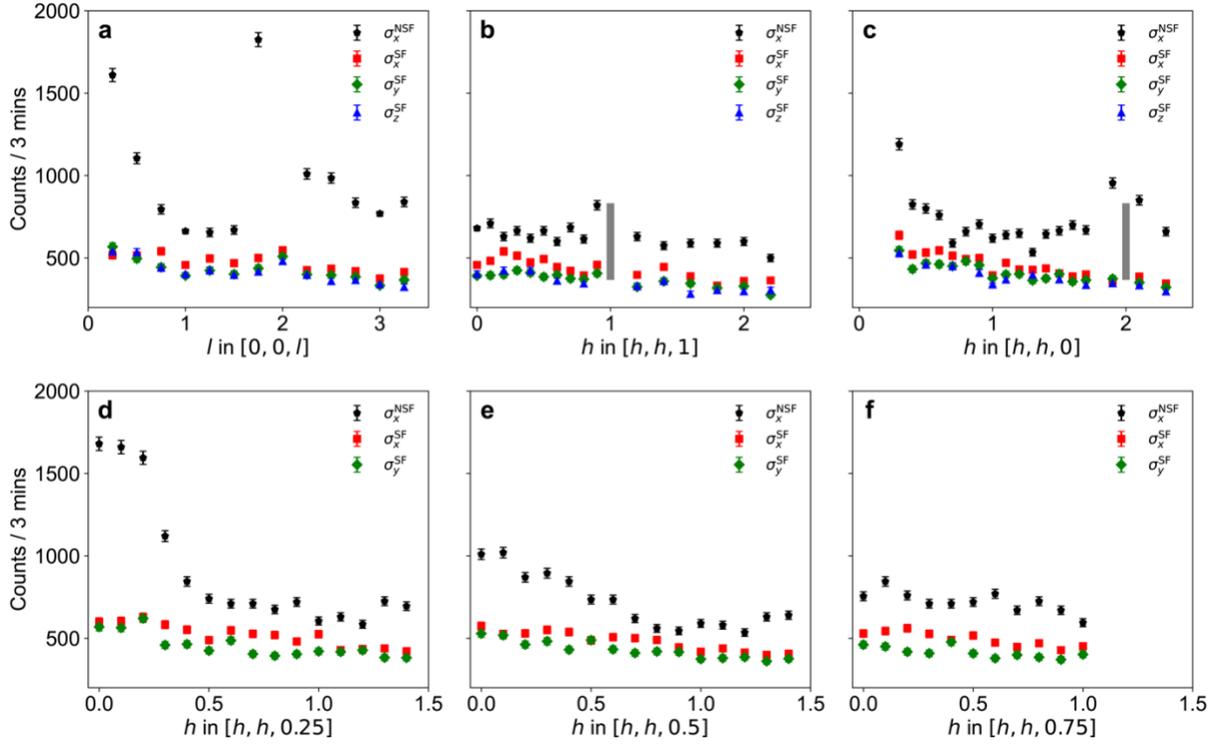

**Extended Figure 3**. **a-f**, Comparison of the polarized NSF neutron scattering cross sections $\sigma_x^{NSF}(Q, E)$ and SF neutron scattering cross sections $\sigma_{x,y,z}^{SF}(Q, E)$ (**a-c**) and $\sigma_{x,y}^{SF}(Q, E)$ (**d-f**) at $E = 0 \pm 0.03$ meV along the [0, 0, $l$], [$h$, $h$, 1], [$h$, $h$, 0], [$h$, $h$, 0.25], [$h$, $h$, 0.5] and [$h$, $h$, 0.75] directions. $\sigma_x^{NSF}(Q, E) > \sigma_{x,y,z}^{SF}(Q, E)$ at all $Q$ points in the scattering plane. Gray windows in panels **b** and **c** indicate nuclear Bragg peaks at (1, 1, 1) and (2, 2, 0) points, respectively. Data are obtained with $E_f = 3.23$ meV. The vertical error bars in **a**–**f** are statistical errors of 1 standard deviation.

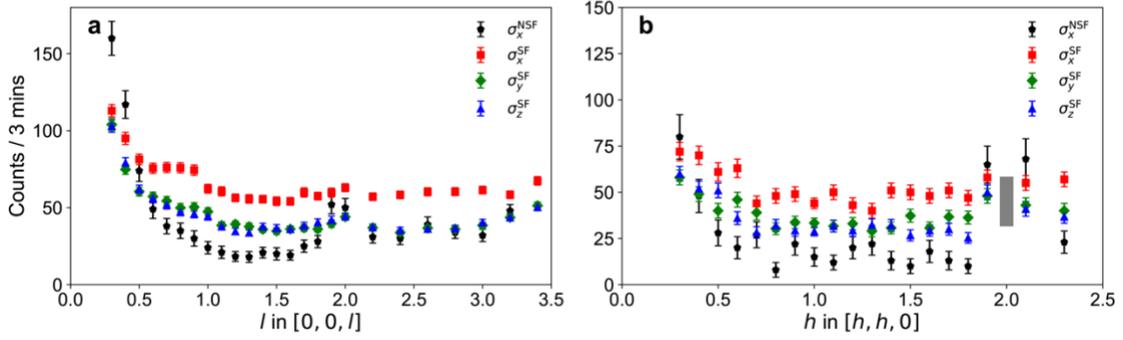

**Extended Figure 4. a-b**, Comparison of the polarized $\sigma_x^{NSF}(\boldsymbol{Q}, E)$ and $\sigma_{x,y,z}^{SF}(\boldsymbol{Q}, E)$ at $E = 0.1 \pm 0.03$ meV along the $[0, 0, l]$ and $[h, h, 0]$ directions. $\sigma_x^{NSF}(\boldsymbol{Q}, E) < \sigma_{x,y,z}^{SF}(\boldsymbol{Q}, E)$ at most $\boldsymbol{Q}$ points in the scattering plane. Gray windows in panel **b** indicate nuclear Bragg peaks at (2, 2, 0) point. The vertical error bars in **a,b** are statistical errors of 1 standard deviation.

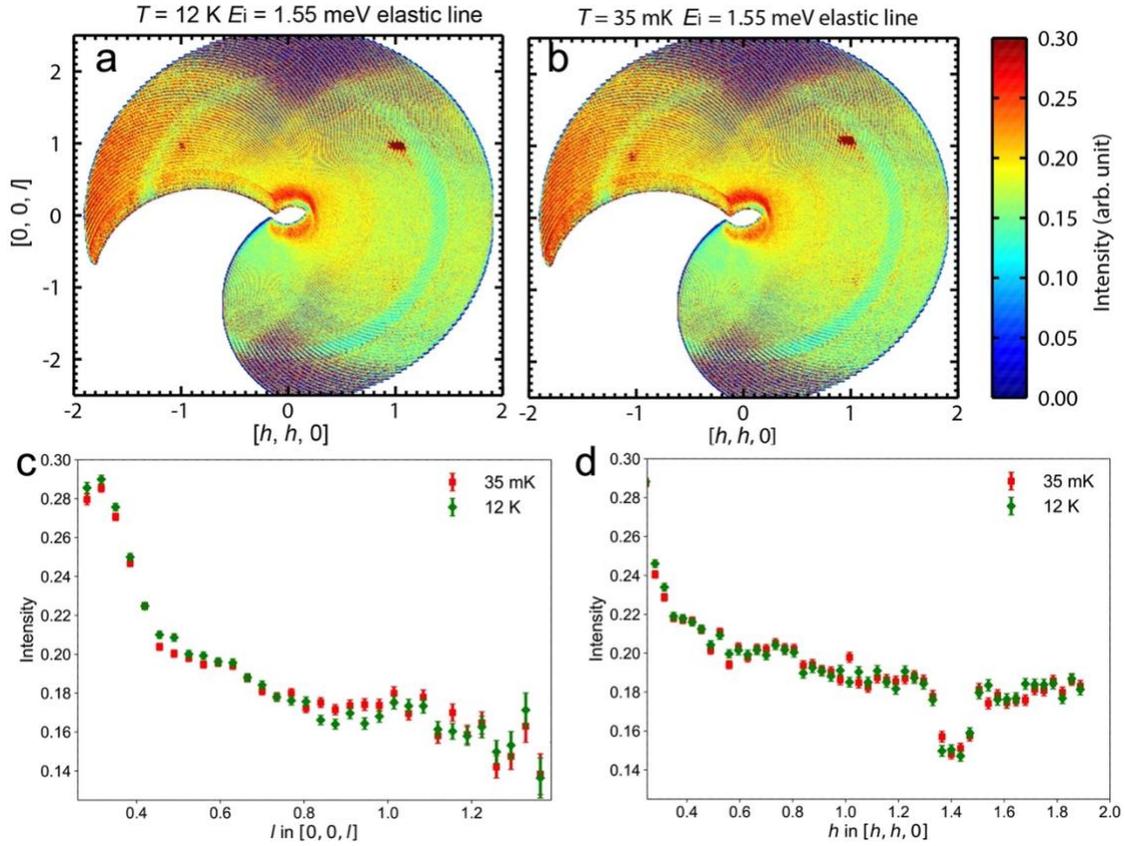

**Extended Figure 5**. **a**,**b**, The raw scattering intensity at 35 mK and 12 K using $E_i = 1.55$ meV at the elastic line ($E = 0 \pm 0.03$ meV) from our previous unpolarized neutron scattering experiment[17]. **c**,**d**, The comparison of raw scattering intensity at 35 mK and 12 K along the [0, 0, $l$] and [h, h, 0] directions from cuts using panel **a**. As one can see, the scattering is highly structured and the scattering has higher intensity at 12 K at almost all $Q$ space probed. The vertical error bars in **c**–**d** are statistical errors of 1 standard deviation.

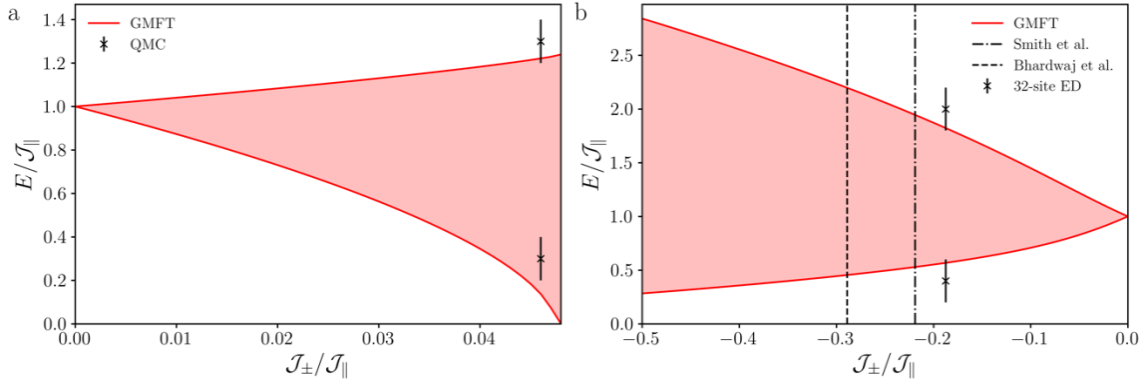

**Extended Figure 6.** Predictions from GMFT for the width of the two-spinon continuum as a function of transverse coupling for **a,** 0-flux QSI and **b,** π-flux QSI. We compare these with QMC results of Ref.[46] and the 32-site ED results of Ref.[28] extracted from the transverse dynamical spin structure factor $S^{\pm}(Q, E)$ for 0- and π-flux QSI, respectively. The dashed and dashed-dotted lines denote the parameter sets obtained in Refs.[27] and [20].

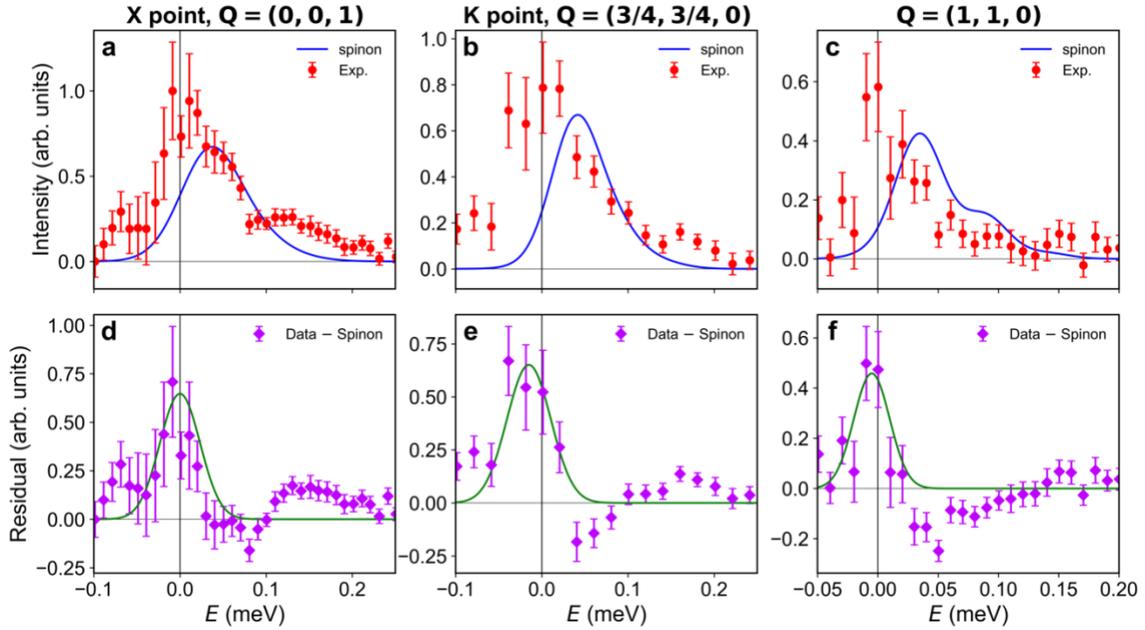

**Extended Figure 7**. **a-c,** Total magnetic scattering $M_z + M_y$ as a function of energy and theoretical prediction for the spinon contribution using $\mathcal{J}_{\parallel} = 0.06$ meV and $\mathcal{J}_{\pm}/\mathcal{J}_{\parallel} = -0.35$ at $Q = (0,0,1)$ (X point), $Q = (3/4,3/4,0)$ (K point), and $Q = (1,1,0)$. The theoretical results are broadened using a Gaussian with a FWHM of 0.076 meV, 0.062 meV, and 0.042 meV at the X, K and $Q = (1,1,0)$ point, respectively. **d-f,** Residual of the fit using only the spinons. The residual is fitted at all three momentum transfers using a Gaussian function centered close to the elastic line.